\documentclass[aps,pre,twocolumn,showpacs,floatfix]{revtex4}
\usepackage{amsmath,bm,epsfig}
\usepackage{latexsym}
\usepackage{amssymb}
\usepackage{color}
\usepackage{morefloats}
\usepackage{subfigure}
\usepackage{graphicx}
\usepackage{amsmath}
\usepackage{bm}
\usepackage{float}
\usepackage{makeidx}
\newcommand{\blue}{\color{black}}
\newcommand{\BLUE}{\color{black}}
\newcommand{\Blue}{\color{black}}

\begin{document}

\title{Revealing directed effective connectivity of cortical neuronal networks from measurements}

\author{Chumin Sun$^1$} \author{K.C. Lin$^1$}
\author{C.Y. Yeung$^1$}
\author{Emily S.C. Ching$^1$\footnote{Corresponding author;
email address: ching@phy.cuhk.edu.hk}} \affiliation {$^1$
Institute of Theoretical Physics and Department of Physics, The
Chinese University of Hong Kong, Shatin, Hong Kong}
\author{Yu-Ting Huang$^{2,3}$} \author{Pik-Yin Lai$^2$}
 \author{C.K. Chan$^3$}
 \affiliation{$^2$Dept. of Physics and Center for Complex Systems,
 National Central University, Chungli, Taiwan 320, ROC}
 \affiliation{$^3$Institute of Physics, Academia Sinica, Taipei, Taiwan 115, ROC}

\date{\today}

\begin{abstract}
In the study of biological networks, one of the major challenges
is to understand the relationships between network structure and
dynamics. In this paper, we model in vitro cortical neuronal
cultures as stochastic dynamical systems and apply
 a method that reconstructs directed networks from dynamics
[Ching and Tam, Phys. Rev. E {\bf 95}, 010301(R), 2017] to reveal
directed effective connectivity, namely the directed links and
synaptic weights, of the neuronal cultures from voltage
measurements recorded by a multielectrode array. The effective
connectivity so obtained reproduces several features of cortical
regions in rats and monkeys and {\Blue has similar network
properties as} the synaptic network of the nematode C. elegans,
the only organism whose entire nervous system has been mapped out
as of today. The distribution of the incoming degree is bimodal
and the distributions of the average incoming and outgoing
synaptic strength are non-Gaussian with long tails. The effective
connectivity captures different information from the commonly
studied functional connectivity, estimated using statistical
correlation between spiking activities. The average synaptic
strengths of excitatory incoming and outgoing links are found to
increase with the spiking activity in the estimated effective
connectivity but not in the functional connectivity estimated
using the same sets of voltage measurements. These results thus
demonstrate that the reconstructed effective connectivity can
{\Blue capture the general properties of synaptic connections and}
better reveal relationships between network structure and
dynamics.
\end{abstract}

\maketitle

\section{Introduction}\label{intro}

The study of networks~\cite{Strogatz,AlbertBarabasi,Newman} has
emerged in many branches of science. Many systems of interest
consist of a large number of components that interact with each
other and can be described as complex networks with the individual
components being the nodes and the interactions among the nodes
represented by links joining the nodes. Understanding how network
structure, which depicts the connectivity or linkage of nodes, is
related to dynamics and to function is a great challenge in
neuroscience~\cite{Bullmore2009,Bassett2017} and in biology in
general. {\blue In neuroscience, three types of connectivity have
been discussed: structural, functional and effective connectivity.
Structural connectivity is the set of physical or anatomical
connections linking neural elements and can be obtained only by
direct measurements, functional connectivity is defined by
statistical dependencies among measurements of neuronal
activities, and effective connectivity refers to the causal
influences exerted by one neural element on another (see
e.g.~\cite{Friston2011,Sporns2011}). As statistical dependency can
arise from indirect interactions, functional connectivity does not
necessarily relate to effective connectivity. Effective
connectivity depends on structure connectivity in that a neural
element can exert causal influences on another neural element only
if the former is linked to the latter by anatomical connections
but structural connectivity itself does not imply effective
connectivity since it is possible that some physical connections
are not being utilized in certain neuronal activities. }

Neuronal cultures grown in vitro serve as a simple but yet useful
experimental model system for studying the relationships between
network structure and the rich dynamics
observed~\cite{Eckmann2007,Feinerman2008,Soriano2015}. One common
technique used to record the activity of neurons in a culture is
the measurement of the {\blue voltage} signals generated by
neurons using a multielectrode array (MEA)~\cite{Obien2015}.
 Estimating connectivity of neuronal
cultures from MEA recordings is thus a problem of great interest.
 Existing methods focus on estimating functional connectivity of
neuronal culture using statistical
correlation~\cite{Maccione2012,Poli2015,Pastore2018} or mutual
information~\cite{Bettencourt2007,Bettencourt2008} of detected
spikes in the MEA recordings.
 {\blue However, one would expect effective
connectivity that gives direct causal influences or interactions
to be more relevant for studying the relationships between network
structure and dynamics and between network structure and
function.}

 {\blue The general problem of inferring networks from dynamics for networked systems whose interacting dynamics are
described by systems of coupled differential equations is a
problem of longstanding interest~\cite{TimmeReview}. It has been
known that statistical correlation often fails to be a good
indicator for direct interactions. There are analytical results
showing that the statistical} covariance of nodal dynamics alone
does not carry sufficient information to recover networks with
directional coupling~\cite{Gilson2016,PhysicaA2018}. For a certain
class of undirected networked systems with bidirectional coupling,
it has been derived that effective connectivity or the information
of direct interactions is contained in the inverse of the
covariance matrix and not the covariance matrix
itself~\cite{noisePRL,PRE,PRErapid}. This result can therefore
explain the finding that a neural population can be strongly
coupled but have weak pairwise correlation~\cite{weak}. {\blue A
number of methods inferring direct interactions from dynamics have
been developed. Some methods assume the dynamics to be linear and
the network to be sparse~\cite{YeungPNAS,SauerPRE}. Others require
additional knowledge such as the functional form of the
dynamics~\cite{Yu,Shandilya,Levnajic,Consensus,SciRep2014} and the
response dynamics of the systems to specific external inputs or
perturbations~\cite{Collins2003,Friston2003,Timme2007}. A
model-independent method has been developed that gives the links
or interactions but not their strength~\cite{Casadiego2017}.}  A
noise-induced relation between the time-lagged covariance and the
equal-time covariance of the dynamics has been derived for {\blue
directed networked systems with linear
dynamics~\cite{Gustavo,MasterThesis} as well as directed networked
systems with nonlinear dynamics around a noise-free steady
state~\cite{ChingTamPRE2017,Lai2017}. Using this covariance
relation for systems with linear dynamics, it has been shown that
the links, except for their directions, can be fully reconstructed
from statistical correlation in the weak coupling
limit~\cite{Timme2017}. Based on this covariance relation, a
method that reconstructs
 directed links as well as the relative strength of the
interactions from dynamics has been
proposed~\cite{ChingTamPRE2017} and validated using numerical
simulations, {\blue not only for systems having stationary
dynamics around a noise-free steady state but also for some
systems that do
not}~\cite{PhysicaA2018,ChingTamPRE2017,TamPhDthesis}.}

In this paper, {\blue we model in vitro neuronal culture as a
stochastic dynamical system and apply this covariance-relation
based method, with suitable modifications, to estimate the
effective connectivity, namely the direct interactions and their
synaptic weights, from the measured voltage signals. We will show
that the effective connectivity reproduces several reported
features of cortical regions in rats and monkeys and {\Blue shares
similar network properties with} the synaptic network of the
nematode C. elegans, the only organism whose entire nervous system
has been mapped out as of today. Moreover, our results will show
that the effective connectivity captures different information
than the functional connectivity based on statistical correlation
of spiking activities estimated from the same sets of measurements
and can better reveal relationships between network structure and
dynamics.}

\section{Data and Method}
\label{method}

\subsection{Experimental measurements}

\begin{figure}[htbp]
\centering
\includegraphics[width=2.5cm,height=2.5cm]{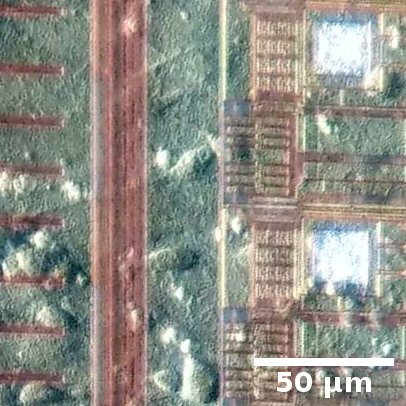} \caption{Surface of high-density
CMOS MEA. The line shows a scale of 50$\mu$m.}
                    \label{fig0}
\end{figure}

Tissues were dissected from {\blue 3} rats and digested with 0.125
\% trypsin for 15 min at 37$^\circ$C {\blue to form a cell
suspension}. A small drop (100 $ \mu$l) of the cell suspension,
containing about $6\times10^{4} $ cells, was plated on the  6~mm
$\times$ 6~mm working area of the complementary
metal-oxide-semiconductor (CMOS)-based high density multielectrode
array (HD-MEA)~(see Fig~\ref{fig0}). After plating on the HD-MEA
chip, cultures were filled with 1~ml of culture medium and placed
in a humidified incubator (5\% CO$_{2}$, 37$^{\circ}$C).

The HD MEA probe (HD-MEA Arena, 3Brain AG) has 4096 electrodes,
which are arranged in a 64 by 64 square grid. The size of each
square electrode is 21 $\mu$m by 21$\mu$m and the electrode pitch
is 42 $\mu$m which gives an active electrode area of 2.67~mm by
2.67~mm. {\blue Spontaneous neuronal activities} were recorded
with the recording device (BioCAM, 3Brain AG) and the associate
software (BrainWave 2.0, 3Brain AG) at 7.06 kHz. One electrode was
used for calibration purpose so there were 4095 electrodes that
recorded 4095 time series of voltage signals. Samples were placed
into the recording device 10 min before the recording in order to
prevent the effects of vibration. Each experimental session lasted
for 5 min and was recorded in dark since the CMOS is a light
active material. {\blue Additional experimental details are
presented in the Appendix.}

\subsection{Method of reconstruction of effective connectivity}

{\blue We estimate the directed effective connectivity for 8 cases
using
 MEA voltage recordings taken at 8 different Days in Vitro (DIV).
In each case, we treat the voltage signal measured by each
electrode after noise reduction by a moving average filter (see
below)} as the activity $x_i(t)$ of node $i$, $i=1,2, \ldots,
4095$, of a neuronal network~\cite{Maccione2012} and model the
dynamics of the network by a generic system of stochastic
differential equations \begin{equation} \frac{d {\bf x}}{dt} =
{\bf F}({\bf x}) + {\bf \eta} \label{generalmodel}
\end{equation}
where ${\bf x} = (x_1, x_2, \ldots, x_N)$ with $N=4095$, {\blue
${\bf F}$ is a general differentiable vector function} and $\eta$
is a Gaussian white noise of zero mean and
$\overline{\eta_i(t)\eta_j(t')} = D_{ij} \delta(t-t')$ that mimics
external influences. The overbar denotes an ensemble average over
different realizations of the noise. {\Blue Assuming small
fluctuations around the asymptotic noise-free solution ${\bf
x}^*$}, we can linearize the equations to obtain
\begin{equation} \frac{d \delta x_i}{dt} = \sum_{j} w_{ij} \delta
x_j
 + \eta_i ,\ \label{network}
\end{equation}
where $\delta x_i(t)=x_i(t)-x^*_i$ and $w_{ij} \equiv {\partial
F_i}/{\partial x_j}({\bf x}^*)$ are the elements of ${\bf W}$,
{\blue which is the Jacobian matrix of ${\bf F}$}.
 When the activity $x_j$ affects {\blue the time evolution of} the
activity
 $x_i$, $w_{ij}$ is nonzero and {\blue this interaction is represented by a link from node $j$ to node $i$ with a weight $w_{ij}$}; otherwise $w_{ij}=0$ and
 there does not exist a link from node $j$ to node $i$. {\BLUE The HD MEA has a high spatial resolution with the size of each electrode being comparable to the size
 of a neuron and the cell density in the culture was low such that the
 voltage measured by an electrode was dominated by the
 electrical signal of one neuron. Thus we
 assume that the activity of each node represents contribution mainly from one neuron and  the interactions
between nodes are through synapses of neurons, and refer the
weights of the interactions to as synaptic weights.} {\blue The
effective connectivity is the information of these direct
interactions among the
 different nodes, which are given by the off-diagonal elements
 of ${\bf W}$. Our goal is to recover
the off-diagonal elements $w_{ij}$ with $i \ne j$ from
$x_i(t)$'s.}

We define the elements of the time-lagged covariance matrix ${\bf
K}(\tau)$ and the equal-time covariance matrix ${\bf K}(0)$ by
\begin{eqnarray} K_{ij}(\tau) &=& \langle [x_i(t+\tau)-\langle
x_i(t+\tau)\rangle][x_j(t)-\langle x_j(t)\rangle] \rangle \qquad \label{Kmatrix}\\
K_{ij}(0) &=& \langle [x_i(t)-\langle
x_i(t)\rangle][x_j(t)-\langle x_j(t)\rangle] \rangle .
\label{Ktaumatrix}\end{eqnarray} where $\langle \ldots \rangle$
denotes a time average. {\Blue For systems that approach a fixed
point in the noise-free limit, ${\bf W}$ is time-independent. Then
by} solving Eq.~(\ref{network}), one
obtains~\cite{Gilson2016,ChingTamPRE2017,Gustavo,MasterThesis}
\begin{equation}
{\bf K}(\tau) = \exp({\tau {\bf W}}) {\bf K}(0) \label{relation}
\end{equation}
{\Blue which implies \begin{equation} M_{ij} = w_{ij}, \qquad {i
\ne j} \label{relation2}
\end{equation}
where \begin{equation} {\bf M} \equiv \frac{1}{\tau} \log ({\bf
K}(\tau) {\bf K}(0)^{-1}) \label{defM}
\end{equation}}
as long as $\tau$ is not too large~\cite{ChingTamPRE2017}. Here,
 $\log$ is the principal matrix logarithm. {\Blue Equation (\ref{relation})
 relates the time-lagged and equal-time covariances ${\bf K}(\tau)$ and ${\bf K}(0)$, which can be calculated using solely the measurements $x_i(t)$'s,
 to ${\bf W}$, which contains information of the direct interactions. The
 importance of Eq.~(\ref{relation2}), which follows from Eq.~(\ref{relation}),} is that the off-diagonal elements
 $M_{ij}$, $i \ne j$, being approximately equal to $w_{ij}$, should separate into two
groups corresponding to $w_{ij}=0$ (no links from node $j$ to node
$i$) and $w_{ij}\ne 0$ (links from node $j$ to node $i$ with
weights $w_{ij}$). Hence for each node $j$, we can infer $w_{ij}$
by clustering the values of $M_{ij}$ for $i \ne j$ into two
groups. {\Blue As demonstrated by numerical
simulations~\cite{ChingTamPRE2017,TamPhDthesis}, this
covariance-relation based method can recover directed and weighted
connectivity not only for the general class of systems as
described above but also for some systems that fluctuate around
oscillatory dynamics modeled by the FitzHugh-Nagumo
dynamics~\cite{FHN}, which is commonly used to model neurons or
fluctuate around the chaotic R\"ossler dynamics~\cite{Rossler}.
For these latter systems that do not approach a fixed point in the
noise-free limit, the off-diagonal elements of ${\bf W}$ are
time-independent and numerical results revealed that
Eq.~(\ref{relation2}) still holds approximately~\cite{noteW} even
though Eq.~(\ref{relation}) cannot be derived. Motivated by these
numerical results, we assume Eq.~(\ref{relation2}) to hold for
some effective time-independent $w_{ij}$'s in our model.}

The principal matrix logarithm is very sensitive to noise in
measurements and a complex matrix could be resulted when the
method is applied to reconstruct realistic systems from
experimental measurements. {\blue Indeed a complex ${\bf M}$ was
obtained when the MEA voltage recordings were directly used in the
calculations.} Let us denote the voltage signals recorded by the
electrodes by $y_i(t)$, $i=1, 2, \ldots, 4095$. We only analyze
measurements taken during which all the 4095 electrodes were
recording properly. When we calculated ${\bf K}_y(\tau)$ and ${\bf
K}_y(0)$ directly from $y_i(t)$ [as defined in
Eqs.~(\ref{Kmatrix}) and (\ref{Ktaumatrix}) with $x_i$ replaced by
$y_i$], we obtained a complex matrix $\log ({\bf K}_y(\tau) {\bf
K}_y(0)^{-1})$. Similar problem has also been reported in a study
of effective connectivity of a cortical network of 68 regions from
fMRI recordings and motivated the development of a Lyapunov
optimization procedure~\cite{Gilson2016}. {\blue In our study, we
are able to avoid this problem by first applying a moving average
filter to the voltage signals to reduce the effect of measurement
noise.} Specifically, we take $x_i(t)=[y_i(t)+y_i(t+\Delta)]/2$,
where $\Delta=0.142$~ms is the sampling time interval and
calculate ${\bf K}(\tau)$ and ${\bf K}(0)$ using $x_i(t)$ with
$\tau=\Delta$. The moving average filter is a simple digital
low-pass filter that reduces random noise while retaining sharp
{\blue changes, if any,} in the data~\cite{Smith1999}. The
resulting matrix ${\bf M}$ is now real. We have further studied
the effect of measurement noise by adding a Gaussian noise to data
obtained in numerical simulations of Eq.~(\ref{generalmodel}) with
$F_i=10x_i(1-x_i)+\sum_{j\ne i}
w_{ij}(x_j-x_i)$~\cite{SunPhDthesis}. Our results show that the
matrix ${\bf M}$ calculated using the noisy data becomes complex
when the standard deviation of the added noise exceeds a certain
threshold  and a real ${\bf M}$ is restored after the above moving
average filter has been applied to the noisy
data~\cite{SunPhDthesis}.

\begin{figure}[htbp]
                    \centering
\includegraphics[width=2.0in]{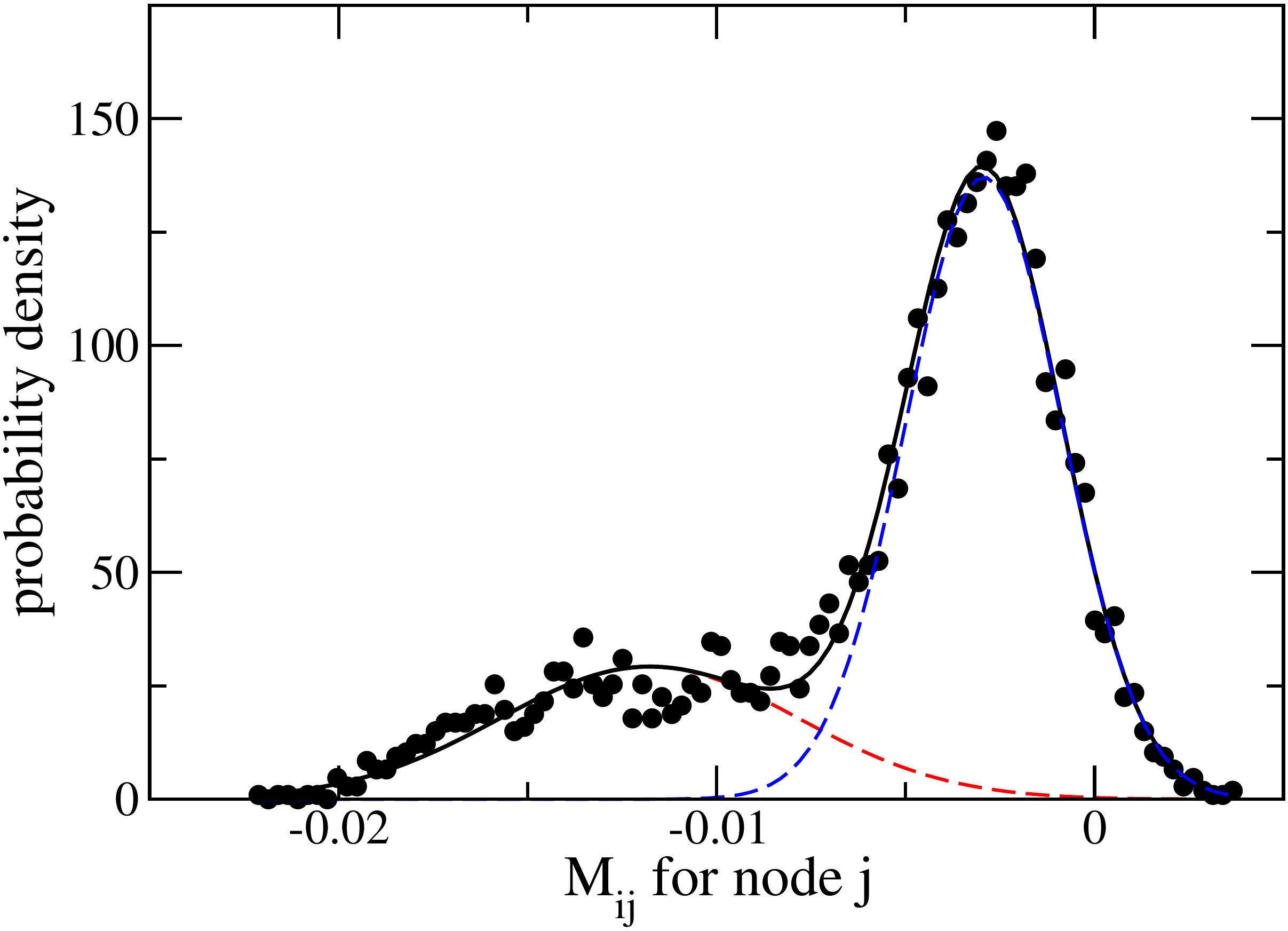} 
\includegraphics[width=2.0in]{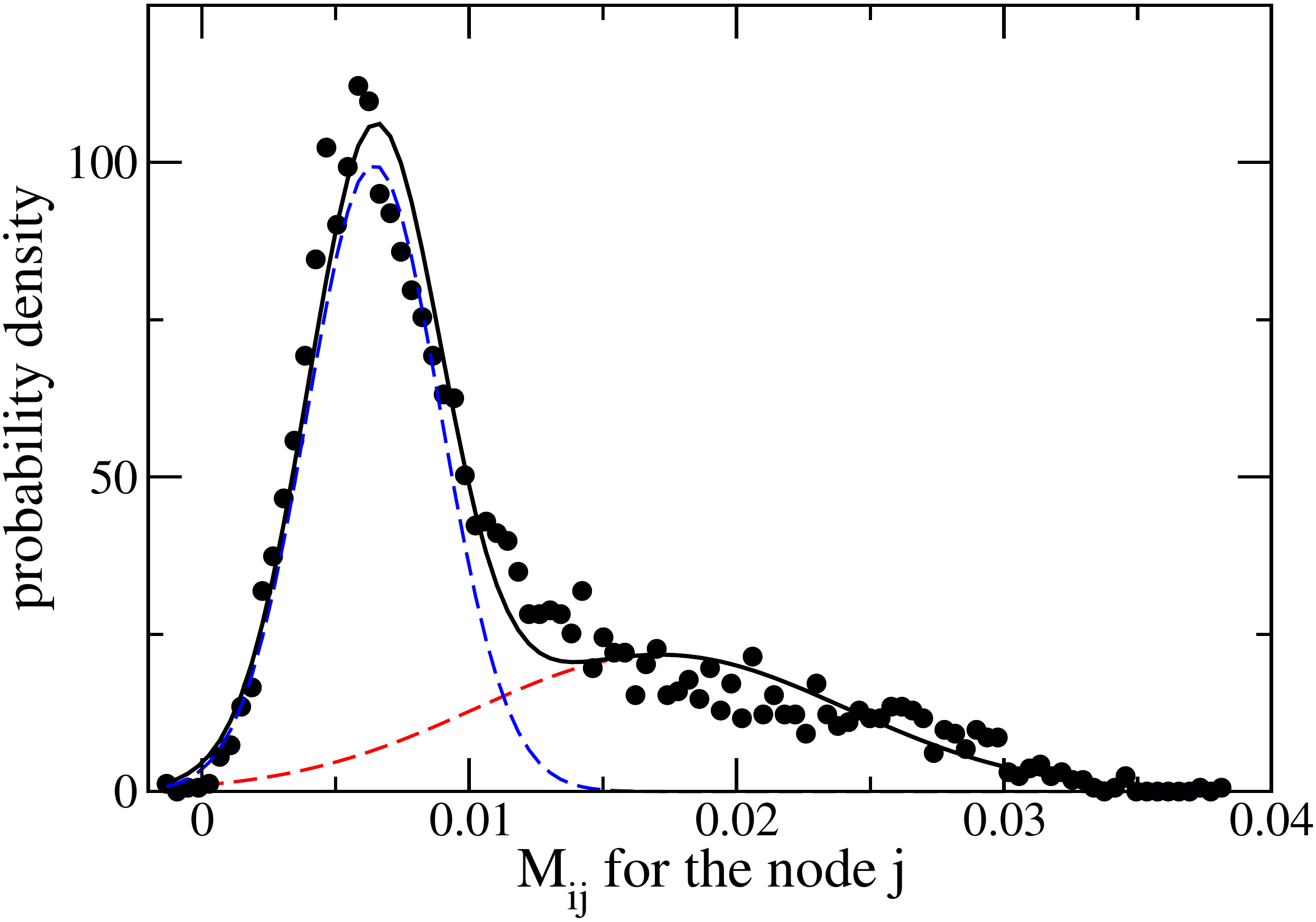}
\caption{The probability density of $M_{ij}$ for node $j$
(circles). {\Blue The top panel shows the result for the 4th node
($j=4$) of case 1 and the bottom panel shows the result for the
4076th node ($j=4076$) of case 8.} The solid black line is the fit
$P_{\rm fit}$ [see Eq.~(\ref{2Gaussianfit})]. The component with a
larger relative proportion with $\alpha$ or $1-\alpha$ greater
than 0.6 (blue dashed line) is identified as the unconnected
component and the other component (red dashed line) is identified
as the connected component. {\Blue In the top panel, the connected
component lies on the left of the unconnected component and the
node $j$ is inferred as an inhibitory node. In the bottom panel,
the connected component lies on the right of the unconnected
component and the node $j$ is inferred as an excitatory node.}}
                    \label{fig1}
\end{figure}

After obtaining the real ${\bf M}$, we extend the clustering
analysis in Ref.~\cite{ChingTamPRE2017}, {\blue with suitable
modifications,} to estimate all the off-diagonal elements
$w_{ij}$'s of the directed effective connectivity matrix as
described below. We assume that the outgoing links of each node,
when exist, can only be all excitatory or all inhibitory. To infer
the outgoing links of a certain node $j$, we fit the distribution
of the values of $M_{ij}$ for all $i \ne j$ by a Gaussian mixture
model of two components: {\blue \begin{eqnarray} \nonumber P_{\rm
fit}(x) = && \alpha \frac{1}{\sqrt{2 \pi} \sigma_1}
\exp\left[-\frac{(x- \mu_1)^2}{2
\sigma_1^2}\right] \\
&& + (1- \alpha) \frac{1}{\sqrt{2 \pi} \sigma_2}
\exp\left[-\frac{(x- \mu_2)^2}{2 \sigma_2^2}\right]
\label{2Gaussianfit}
\end{eqnarray}}
This is done by using MATLAB `fitgmdist', {\blue which is based on
the iterative Expectation-Maximization algorithm~\cite{MATLAB1}.}
Two examples of these fits are shown in Fig~\ref{fig1}. {\blue
According to Eq.~(\ref{relation2}), one expects that the
unconnected component of non-existent links of $w_{ij}=0$ should
have a mean close to 0 but as can be seen in Fig.~\ref{fig1}, the
means of the two fitted Gaussian components both deviate from
zero. The presence of hidden nodes whose signals are missing can
cause a shift in the values of
$M_{ij}$'s~\cite{ChingTamHidden2018} and such hidden nodes could
be neural cells lying outside the active electrode area of the MEA
probe whose voltage signals were not detected~(see Sec. III D).
Thus, to improve the performance of our method in the presence of
hidden nodes, we make use of the sparsity of the network to
identify the unconnected component whenever possible.
Specifically, when the two components are well separated with
$|\mu_1-\mu_2|> \sigma_1+\sigma_2$, we first check whether
$\alpha$ or $(1-\alpha)$ is greater than 0.6 and, if yes, identify
the component of the larger proportion as the unconnected
component corresponding to $w_{ij}=0$. If neither $\alpha$ nor
$1-\alpha$ exceeds 0.6, we then take the component whose mean is
closer to zero as the unconnected component.} The remaining
component is referred as the connected component. If the connected
component is on the right of the unconnected component as shown in
the top panel of Fig~\ref{fig1}, then node $j$ is an excitatory
node with all outgoing links of $w_{ij}>0$. Otherwise if the
connected component is on the left of the unconnected component as
shown in the bottom panel of Fig.~\ref{fig1}, then node $j$ is an
inhibitory node with all outgoing links of $w_{ij}<0$. {\blue For
each node $j$, the data points $M_{ij}$ for $i \ne j$ are
clustered into each of the two components according to the
probability $p_i$ of each of the $M_{ij}$ values belonging to the
unconnected component. We obtain $p_i$'s using  MATLAB `cluster'
which performs agglomerative clustering~\cite{MATLAB2}.} If $p_i
>0.5$, then $w_{ij}=0$ and there is no link from node $j$ to node
 $i$. Otherwise if $p_i\le 0.5$, then there is a link from
node $j$ to node $i$ with a weight $w_{ij} = M_{ij} - \langle
M_{kj} | w_{kj}=0 \rangle_k$ where $\langle M_{kj} | w_{kj}=0
\rangle_k$ is the average over $k$ of those $M_{kj}$ values that
are estimated to correspond to
 $w_{kj}=0$. We repeat this procedure for all the
nodes $j$ to estimate all the off-diagonal elements $w_{ij}$ with
$i \ne j$.

In the event that the two Gaussian components are not well
separated with $|\mu_1-\mu_2| < {\rm max} (\sigma_1,\sigma_2)$, we
fit the distribution of $M_{ij}$ again by one single Gaussian
distribution, denoted by $P_G(x)$ of mean $\mu$. {\blue We denote
the smooth distribution of $M_{ij}$ estimated using MATLAB
`ksdensity', which is based on a normal kernel smoothing function
with an optimal bandwidth~\cite{MATLAB3}, by $P_K(x)$. We identify
the outliers, which are data points whose values of $P_K(x)$
deviate significantly from $P_G$, as the connected component with
$w_{ij}\ne 0$. Precisely, we define $x_I$ and $x_E$ by}
\begin{eqnarray} x_E = \min_{x> \mu} \{
P_K(x)= 3P_G(x)\}\\
x_I = \max_{x < \mu} \{ P_K(x)= 3P_G(x)\} \end{eqnarray} {\blue
then calculate the number of data points in the two groups: (1)
$M_{ij}
> x_E $ and (2) $M_{ij} <
x_I $ and identify the bigger group of these data points as the
connected component with $w_{ij} = M_{ij}- \langle M_{kj} |
w_{kj}=0 \rangle_k$.  If no such outliers exist,} then node $j$ is
inferred to have no detectable outgoing links. For the in-between
cases of the two Gaussian components that are neither well
separated nor too close to each other with ${\rm max}
(\sigma_1,\sigma_2) \le |\mu_1-\mu_2| \le \sigma_1+\sigma_2$, we
choose either the two-Gaussian fit or the single-Gaussian fit
according to the Bayesian information criterion for fitting models
selection~\cite{Kass1995,Raftery1999}.

\section{Results of the Effective Connectivity}
\label{results}

{\blue For each of the 8 cases,} we study the basic network
measures and the distributions of degree and synaptic strength.

\subsection{Basic network measures}
\label{basic}

We calculate several basic network measures including the
connection probability $p$, the ratio $r_B$ of the number of
bidirectionally connected pairs to the expected number for a
random network with same connection probability $p$, the fractions
$f_E$ and $f_I$ of excitatory and inhibitory nodes, the fraction
$f_{\rm SCC}$ of nodes that form the strongly connected component,
the characteristic path length $l$, the average clustering
coefficient (CC) and the small world index (SWI). The results are
shown in Table~\ref{table1}. The connection probability $p$ of a
network of $N$ nodes with $N_L$ links is defined by
$p=N_L/[N(N-1)]$. We find that $p$ ranges from 0.7-1.9\% which is
consistent with our assumption that the neuronal networks are
sparse. This average value of $p$ is smaller but comparable to
that of the chemical synapse network of C.
elegans~\cite{Varshney2011}. Most of the connections are
unidirectional in accord with the directional transmission of
signals in neurons. One expects neuronal networks to be organized
and thus highly nonrandom in order to facilitate effective and
efficient signal transmission. For a random network of $n$ nodes
and connection probability $p$, the expected number of
bidirectionally connected pairs is given by $N(N-1)p^2/2$.  We
denote the ratio of the number of bidirectionally connected pairs
in the network to the expected number in a random network by
$r_B$. The values of $r_B$ exceed 4 for all the networks,
consistent with the well-documented over-representation of
bidirectional connected pairs in rat cortical
circuits~\cite{SongPLoS2005,MarkramJPhysio1997,SjostromNeuron2001,PerinPNAS2011}.

\begin{table*}[htbp]
                \centering
                \setlength{\tabcolsep}{1mm}{
                \begin{tabular}{|c|cccccccc|c|} \hline
 & case 1    & case 2    & case 3    & case 4    & case 5    & case 6    & case 7   & case 8  & C. elegans
 \\\hline
DIV & 11 & 22 & 25 & 33 & 45 & 52 & 59 & 66 & --- \\
$p$ (\%) & 1.2 & 1.9 & 1.4 & 1.5 & 1.1 & 1.7 & 1.4 & 1.5  &  2.8 \\
$r_B$ & 5.9 & 5.5  & 10.7 & 5.7  & 6.5  & 4.21  & 4.0  & 4.4  & --- \\
$f_E$ & 0.62 & 0.80  & 0.84  & 0.75  & 0.62  & 0.66  & 0.48  &
0.57  & --- \\
$f_I$ & 0.27 & 0.13  & 0.14  & 0.18  & 0.21  & 0.24  & 0.32  &
0.28  & --- \\
$f_{\rm SCC}$ & 0.88 & 0.93 & 0.98 & 0.92 & 0.81 & 0.90 & 0.77 & 0.83  & 0.85 \\
$l$ & 4.0  & 3.7  & 3.7  & 3.9  & 4.1  & 3.7  & 3.8  & 3.8    & 3.5  \\
CC & 0.26  & 0.36  & 0.38  & 0.30  & 0.25  & 0.28  & 0.18  & 0.22    & 0.22 \\
SWI & 13.1  & 11.3  & 17.1  & 12.7  & 14.4  & 10.4  & 7.9  & 8.8 &
2.3 \\  \hline
 \end{tabular}
}\caption{Basic network measures of the networks reconstructed.
                When available, the corresponding results for the chemical synapse
network of C. elegans~\cite{Varshney2011} are included for
comparison.} \label{table1}
\end{table*}

Each node is inferred as excitatory or inhibitory according to the
sign of the synaptic weights of its outgoing links as discussed in
Sec.~\ref{method}. There is a small fraction of nodes with no
detectable outgoing links. As can be seen in Table~\ref{table1},
the values of $f_I$ range from 0.13 to 0.31, which are comparable
to the measured fractions (0.15-0.30) of inhibitory neurons in
various cortical regions in monkey~\cite{Hendry1987}. The balance
between excitation and inhibition in the cortex is believed to
play an important role in executing proper brain functions and
disruption of such a balance may underlie the behavioral deficits
that are observed in conditions such as autism and
schizophrenia~\cite{Rubenstein2003,Levitt2004,Yizhar2011,Lewis2011,Marin2012}.
The fraction $f_{\rm SCC}$ of nodes that form the largest strongly
connected component exceeds 70\% for all the cases studied. The
characteristic path length $l$, which is the average shortest path
length for all nodes that can be connected by a finite path is
about 4, and the average local clustering coefficient CC, which is
the average of the connection probability of the outgoing
neighbors of each node, ranges from $0.18-0.38$. These {\Blue
network properties} are comparable to those of the chemical
synapse network of C. elegans. Furthermore, the reconstructed
neuronal networks all have small-world topology, with small-world
index SWI substantially greater than one. It has been shown that
small-world {\BLUE networks allow} efficient transmission of
information~\cite{LatoraPRL2001}.

\subsection{Distributions of incoming and outgoing degrees}
\label{degree}

\begin{figure}[htbp]
                    \centering
\includegraphics[width=2.5in]{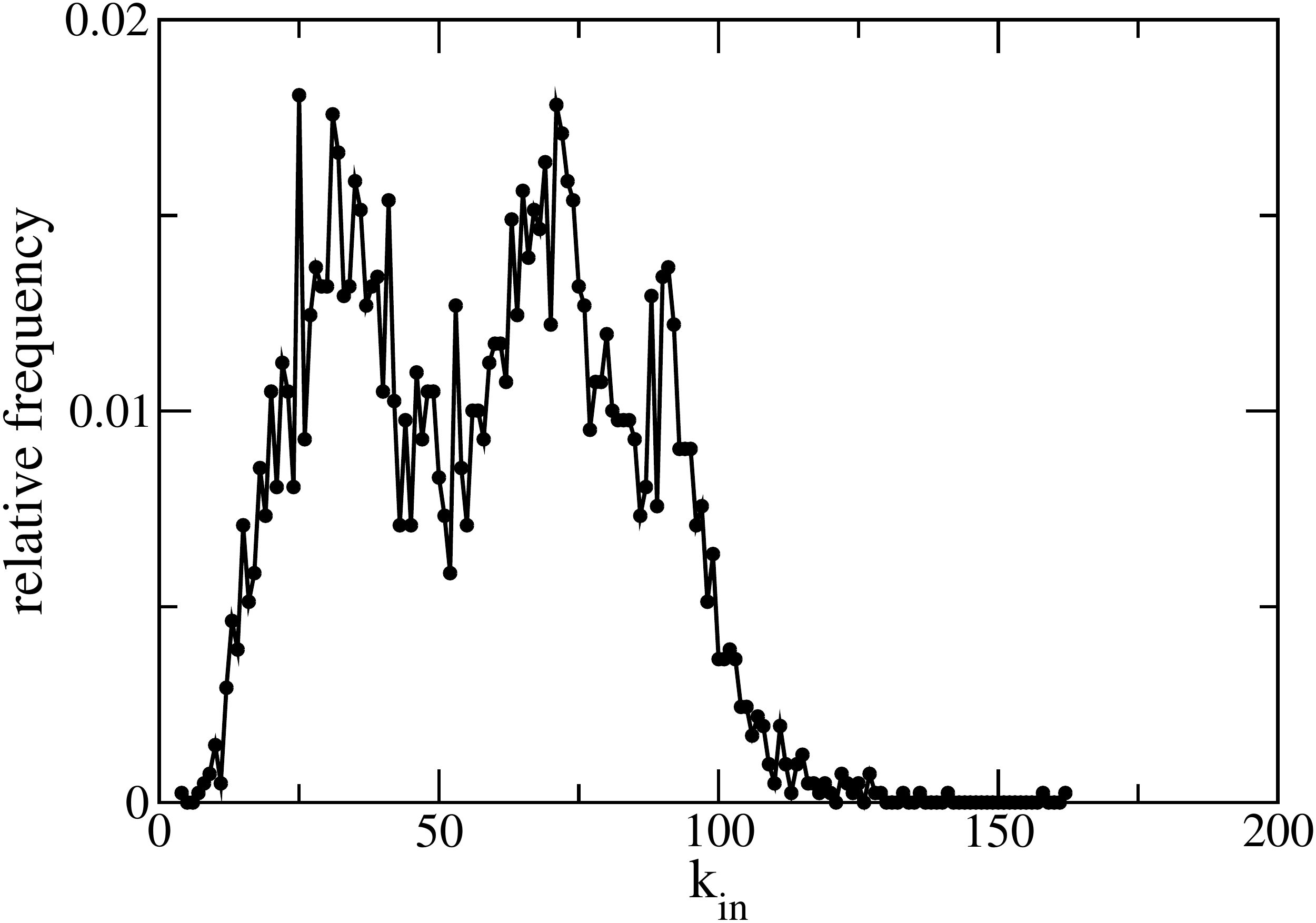}
\includegraphics[width=2.5in]{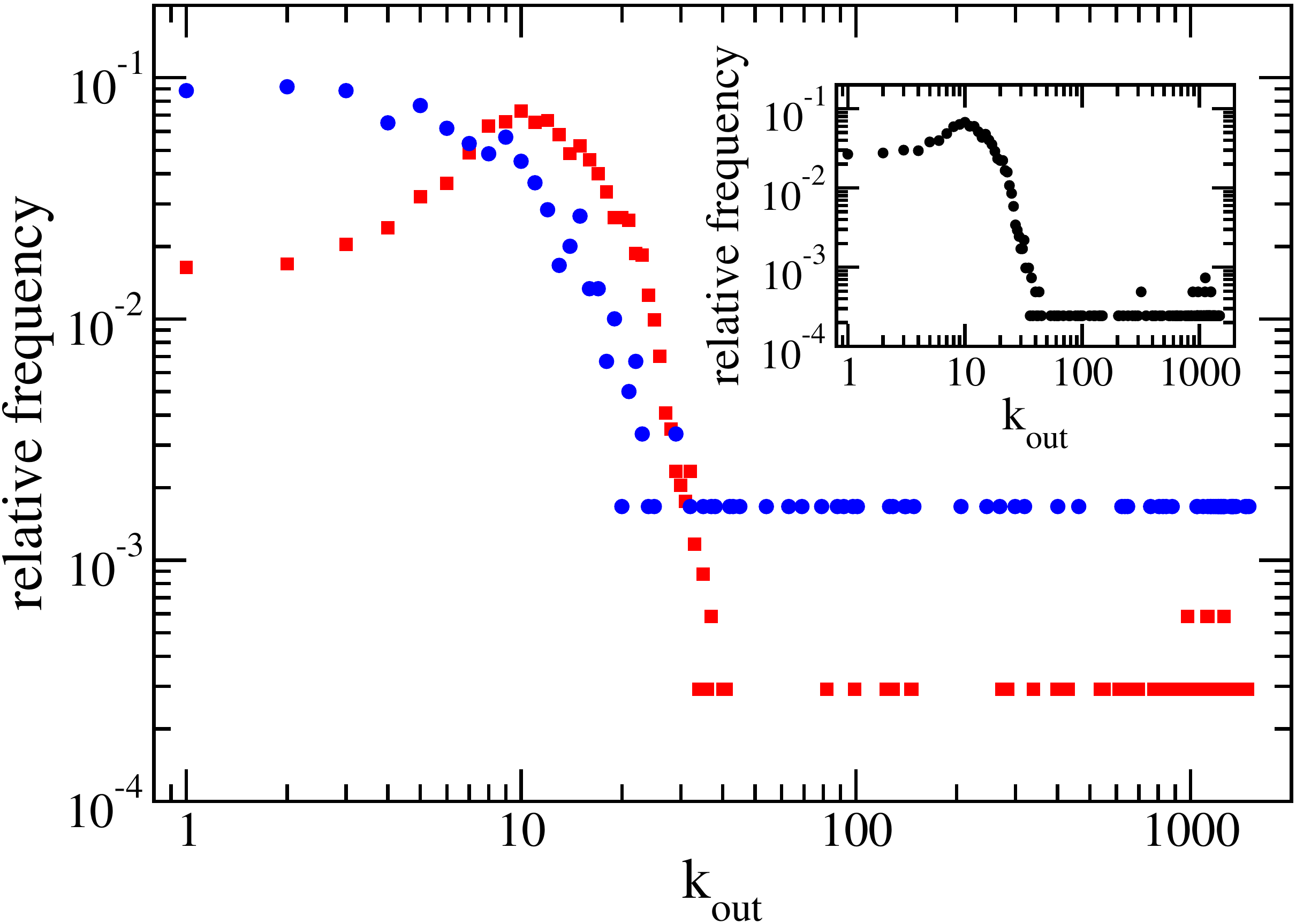}
\caption{Distributions of the incoming degree $k_{\rm in}$ (top
panel) and outgoing degree $k_{\rm out}$ (bottom panel) of case 3.
In the bottom panel, we show the distribution of $k_{\rm out}$
separately for excitatory (circles) and inhibitory nodes (squares)
and for all the nodes in the inset.}
                    \label{fig2}
                    \end{figure}

\begin{figure}[htbp]
                    \centering
\includegraphics[width=2.5in]{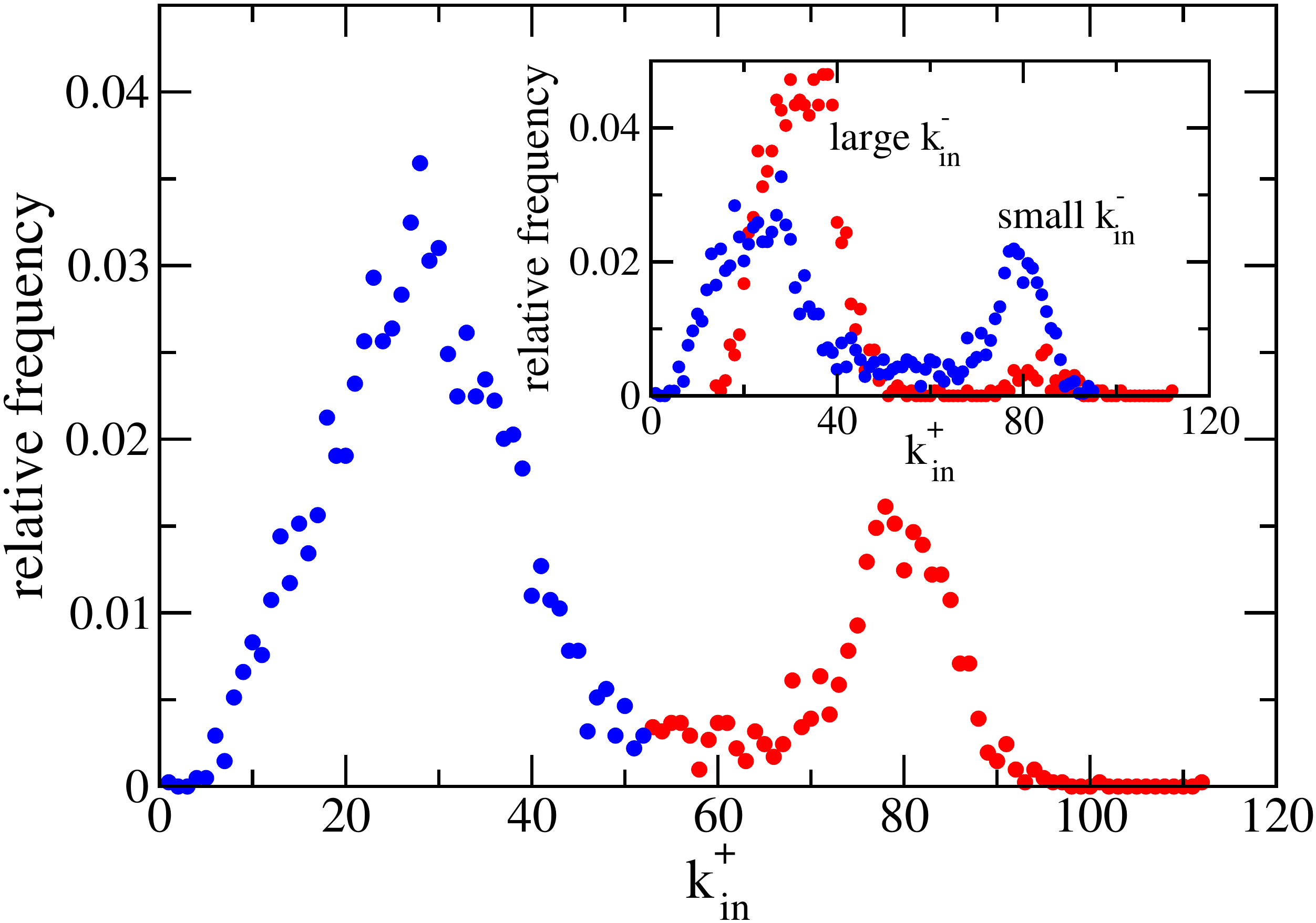}
\includegraphics[width=2.5in]{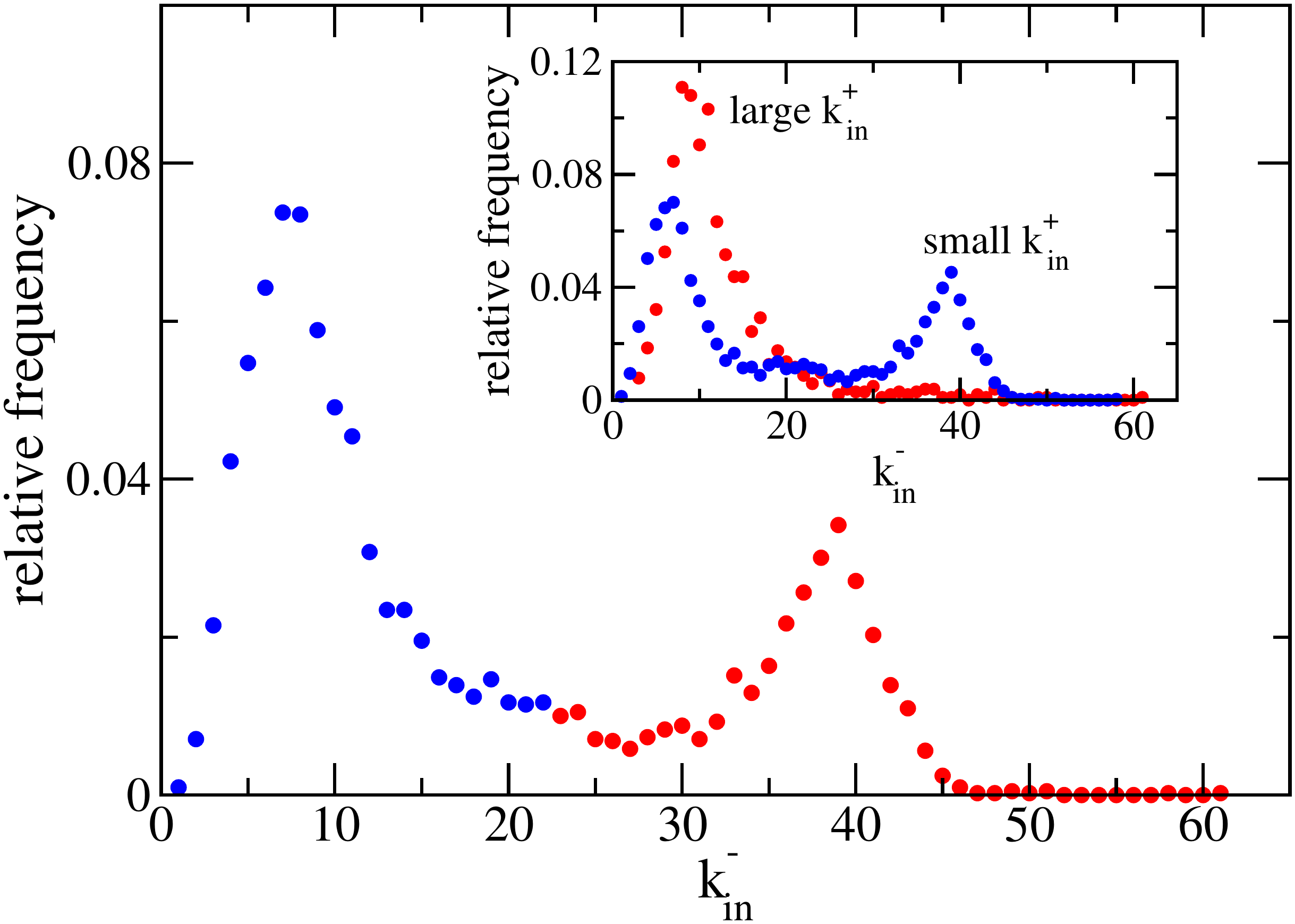}
\caption{Distributions of the excitatory and inhibitory incoming
degrees,
 $k_{\rm in}^+$ (top panel) and $k_{\rm in}^-$ (bottom panel),
of case 3. The distributions are clearly bimodal with two modes of
large and small $k_{\rm in}^+$ or $k_{\rm in}^-$. In the insets,
we show the relative frequencies separately for the two modes of
nodes of small and large $k_{\rm in}^-$ (or $k_{\rm in}^+$). }
                    \label{fig3}
                    \end{figure}

The distributions of the incoming and outgoing degrees are
qualitatively the same for all the 8 cases and results for one
case are shown in Fig.~\ref{fig2}. There are several notable
features. First, most of the nodes have a small $k_{\rm out}$ that
is less than a few tens but a small fraction of nodes have an
exceptionally large $k_{\rm out}$ exceeding a thousand. This is
the case for both the excitatory and inhibitory nodes.
 Second, the incoming degree distribution is approximately
bimodal, which is different from the approximate scale-free
distribution reported for functional
connectivity~\cite{Pastore2018}~(see also Sec.~\ref{comparison}).
The bimodal feature of the incoming degree distribution is more
clearly revealed in the separate distributions of the excitatory
and inhibitory incoming degrees $k_{\rm in}^+$ and $k_{\rm in}^-$
for incoming links of positive and negative synaptic $w_{ij}$,
respectively~(see Fig.~\ref{fig3}). By studying the distribution
of $k_{\rm in}^+$ (or $k_{\rm in}^-$) separately for the two modes
of nodes of small and large $k_{\rm in}^-$ ($k_{\rm in}^+$) (see
inset of Fig.~\ref{fig3}), it can be seen that nodes tend not to
have both large $k_{\rm in}^+$ and large $k_{\rm in}^-$.

\subsection{Distributions of average synaptic strength}
\label{swdist}

\begin{figure}[htbp]
\centering
\includegraphics[width=2.6in]{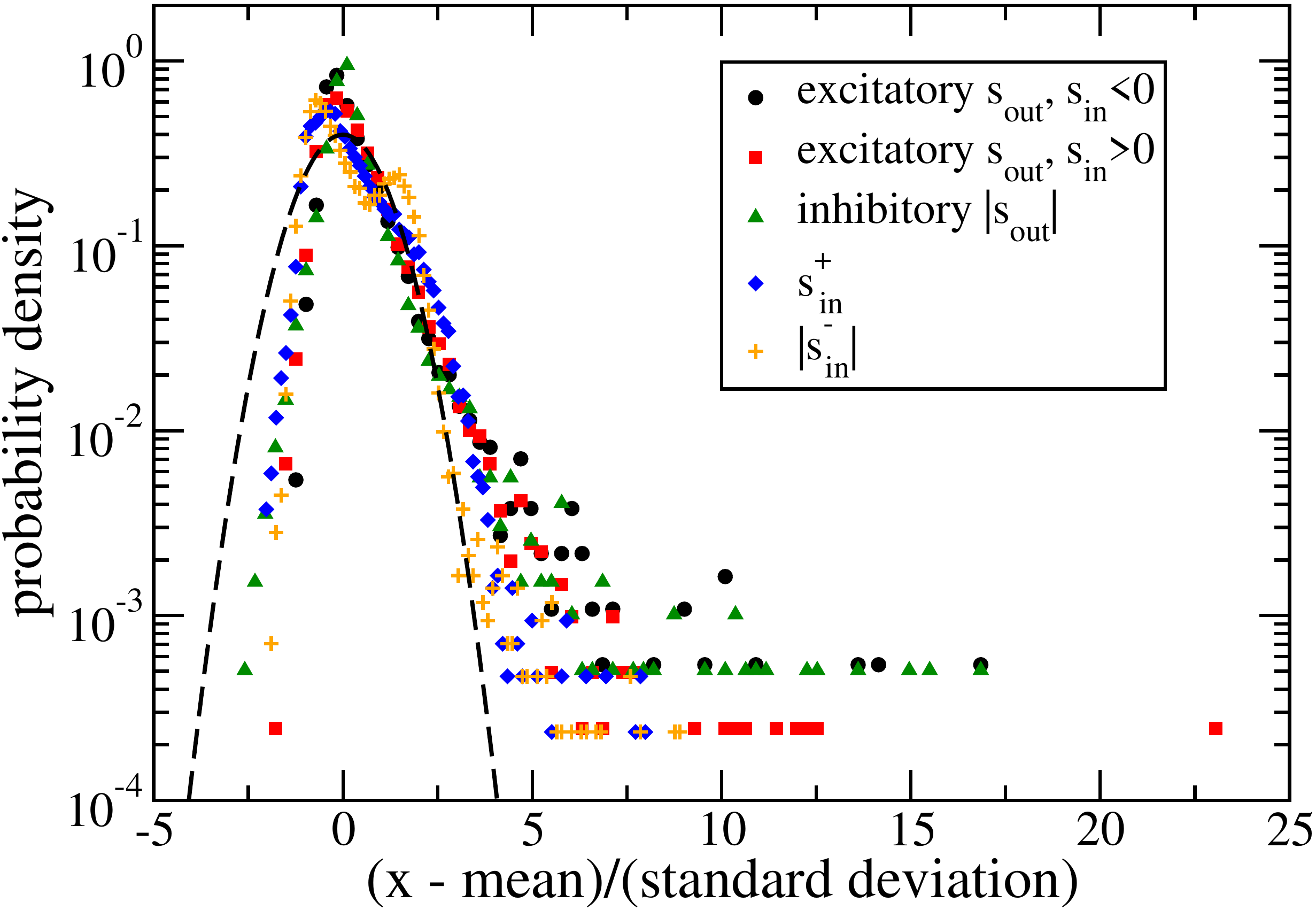} 
\includegraphics[width=2.6in]{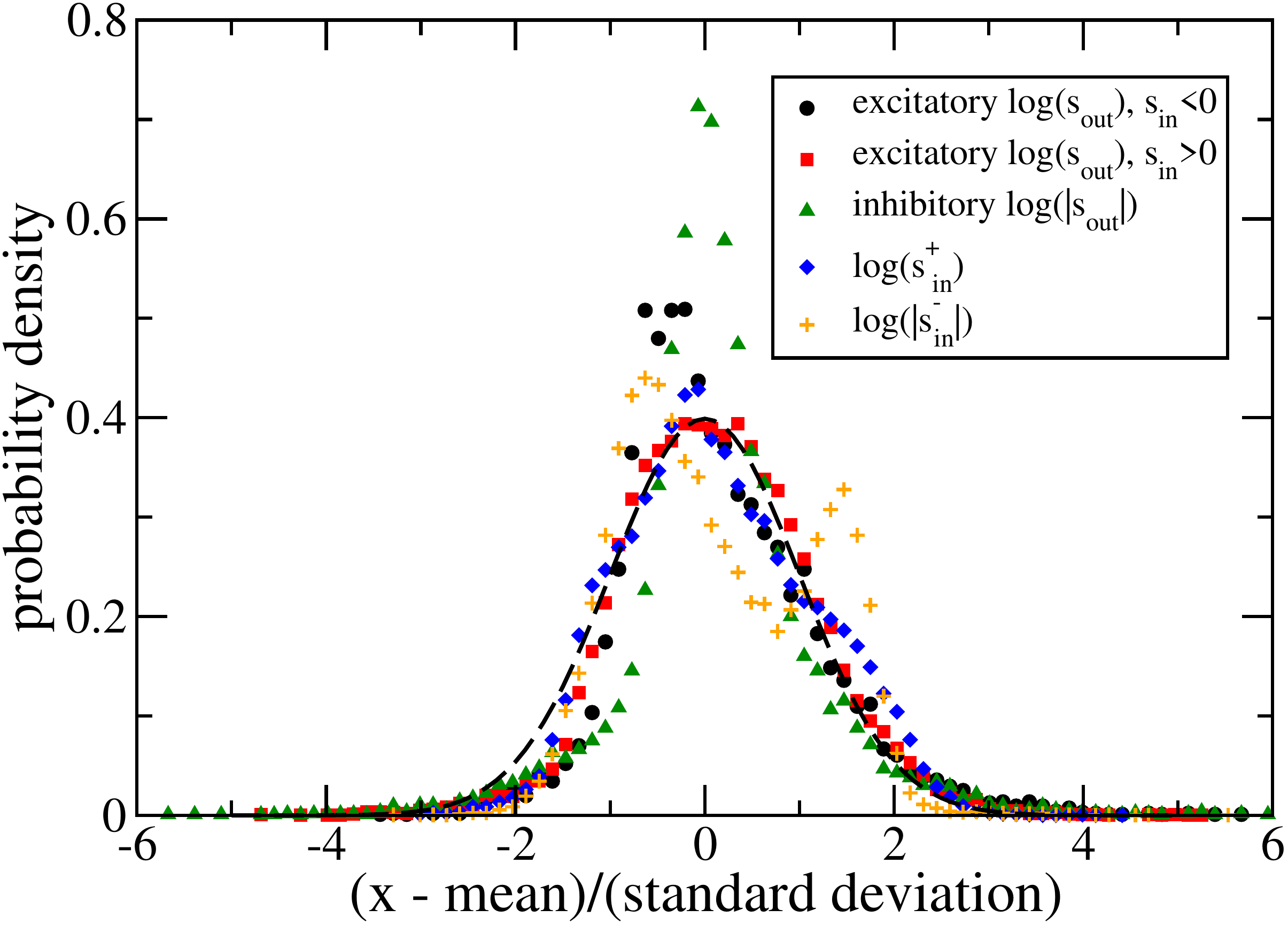}
\caption{Distributions of the average synaptic strengths of
excitatory and inhibitory incoming links $s_{\rm in}^+$ and
$s_{\rm in}^-$ and of outgoing links $s_{\rm out}$. Top panel:
Distributions of the standardized values of the average synaptic
strength, $(x-{\rm mean})/(\mbox{standard deviation})$, where $x$
is $s_{\rm in}^+$, $s_{\rm in}^-$ or $s_{\rm out}$ in each case.
Bottom panel: Distributions of the standardized values of the
logarithm of the average synaptic weights with $x$ now being
$\log(s_{\rm in}^+)$, $\log(|s_{\rm in}^-|)$ or $\log(|s_{\rm
out}|)$. The dashed line is the standardized Gaussian
distribution.} \label{fig4}
\end{figure}

Both in vitro and in vivo studies have indicated that the
distribution of synaptic weights in the cortex are generally
skewed with long tails and typically lognormal~\cite{Buzsaki2014}.
Motivated by these findings, we study the distributions of the
average synaptic strength of the links. The various averages of
the synaptic weights are defined as follows
\begin{eqnarray}
s_{\rm in}(i) &\equiv& \frac{\sum_{j \ne i} w_{ij}}{k_{in}(i)} \label{sin} \\
s_{\rm in}^+(i) &\equiv& \frac{\sum_{j\ne i, w_{ij}>0} w_{ij}}{k_{\rm in}^+(i)} \label{sinplus} \\
 s_{\rm in}^-(i) &\equiv& \frac{\sum_{j\ne i, w_{ij}<0} w_{ij}}{k_{\rm in}^-(i)} \label{sinminus} \\
 s_{\rm out}(i) &\equiv& \frac{\sum_{j\ne i} w_{ji}}{k_{\rm out}(i)} .
 \label{sout}
 \end{eqnarray}
Among these quantities, $s_{\rm in}^+ \ge 0 $ is the average of
the synaptic weights of excitatory incoming links and $s_{\rm
in}^- \le 0$ is the average of the synaptic weights of inhibitory
incoming links, $s_{\rm in}$ is the average of the synaptic
weights of all incoming links and can be positive or negative, and
$s_{\rm out}$ is the average of the synaptic weights of all
outgoing links and is positive for excitatory nodes, negative for
inhibitory nodes and zero for the nodes with no detectable
outgoing links. To have better statistics, we use the data from
all the 8 networks to calculate the distributions. We first
calculate the mean and standard deviation of these average
synaptic weights in each of the networks and then calculate the
distribution of the standardized values, which are the values
subtracted by the mean and divided by the standard deviation. The
distribution of $s_{\rm out}$ for excitatory nodes is found to
depend on whether $s_{\rm in}$ is positive or negative. As seen in
Fig~\ref{fig4}, all these average synaptic strengths have a
non-Gaussian distribution that is skewed with long tails. This
indicates that a small fraction of the nodes have dominantly
strong average synaptic strengths and thus the average synaptic
strength of the links of the nodes are not well represented by the
mean values. We also calculate the distribution of the
standardized values of the logarithm of these average synaptic
strengths and find that $s_{\rm out}$ for excitatory nodes with
$s_{\rm in}>0$ has an approximately lognormal distribution~(see
Fig.~\ref{fig4}).

\subsection{Possible effects of hidden nodes} \label{hidden}

\begin{figure}[htbp]
\centering
\includegraphics[width=2.2in]{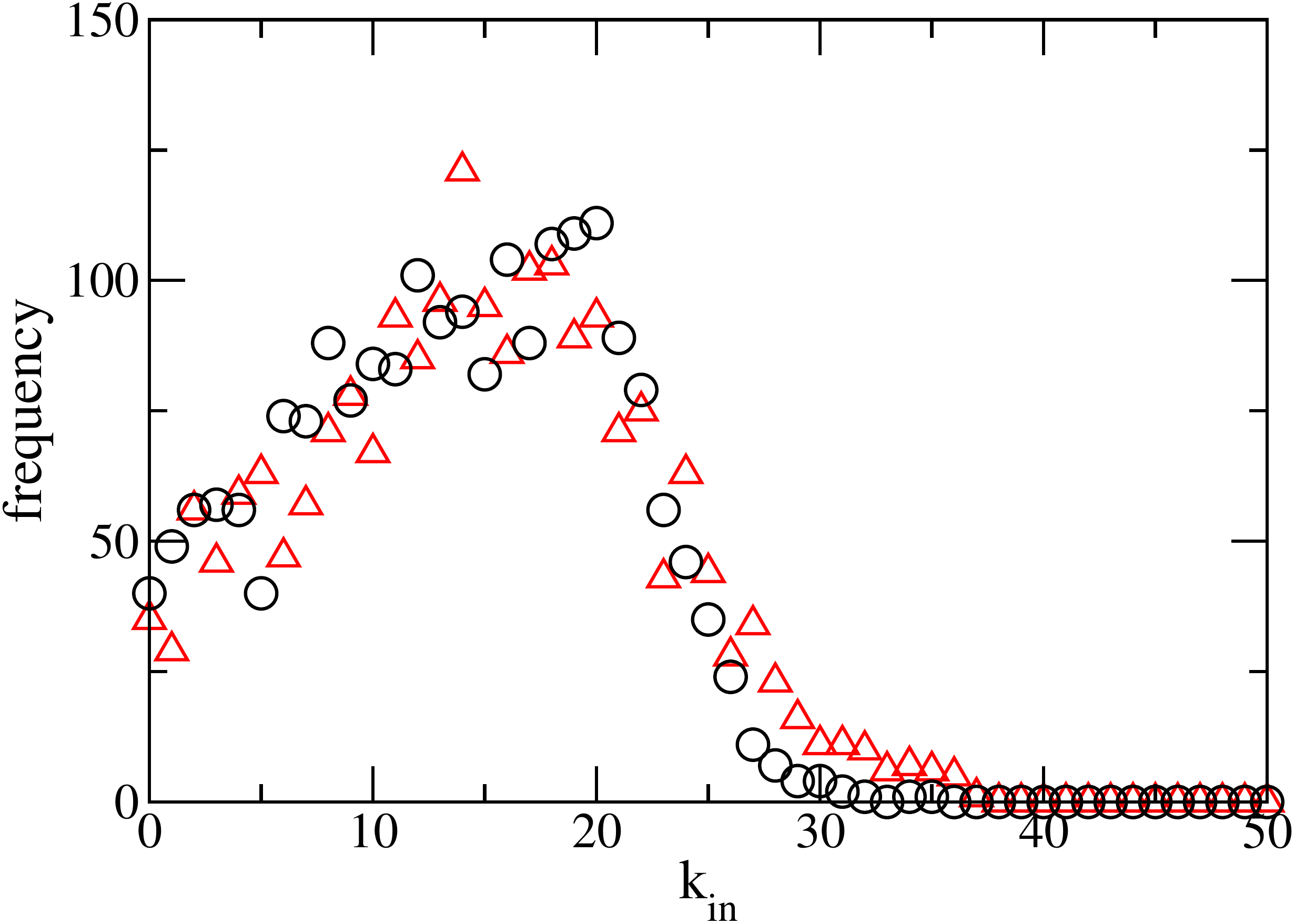}
\includegraphics[width=2.2in]{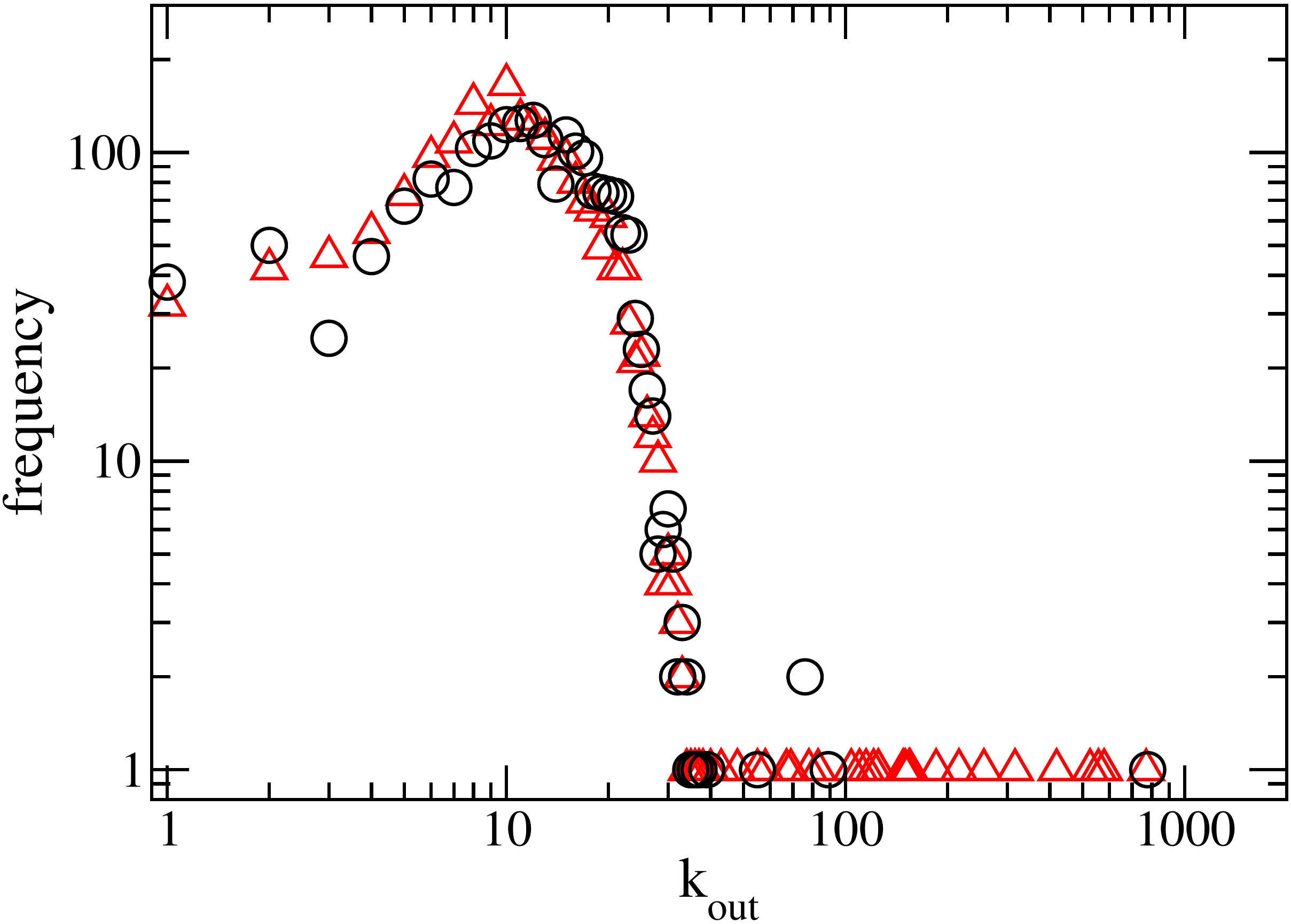}
\caption{Distributions of in- and out-degrees of the subnetwork
(circles) and partial network (triangles) for case 3.}
\label{fig5}
\end{figure}

Neural cells lying in the working area outside the active
electrode area of the MEA probe could form synaptic connections
with cells within the active electrode area but their spontaneous
activities were not recorded. Thus the MEA recordings could miss
out information from those nodes that were not detected. These
nodes are thus the so-called hidden nodes. It has been shown that
in bidirectional networks the presence of hidden nodes that are
randomly missed out has no significant adverse effects on the
reconstruction of the links among the measured
nodes~\cite{ChingTamHidden2018}. To investigate whether and how
undetected signals from the hidden nodes might affect our results,
we reconstruct a partial network using only recordings of 2025
electrodes, which are the 45 by 45 electrodes in the central
region of the active electrode area, and compare it with the
subnetwork of these corresponding 2025 nodes, extracted from the
whole network that we reconstructed using recordings of all the
4095 electrodes for case 3. We find that the partial network
captures correctly 99.8\% of the non-existent links and 83.0\% of
the links in the subnetwork. Moreover, the partial network and the
subnetwork have similar in- and out-degree distributions as shown
in Fig.~\ref{fig5}. These results support that the directed
effective connectivity obtained using the 4095 electrodes is not
significantly affected by the possibly missed out signals from the
neural cells lying outside the active electrode area of the MEA
probe.

\section{Comparison of Effective Connectivity and Functional
Connectivity} \label{comparison}

The effective connectivity studied in the present work is proposed
to capture the direct interactions among the nodes based on the
general class of model described in Eq.~(\ref{generalmodel}) while
functional connectivity studied in existing methods is based
mainly on the statistical dependence of the dynamics of the nodes.
It is thus expected that effective connectivity and functional
connectivity to contain different information about the system. In
this Section, we compare directly the two types of connectivity
estimated from the same sets of measurements. We follow the method
in Ref.~\cite{Pastore2018} to estimate the functional connectivity
using the statistical correlation of the spikes detected from the
MEA recordings. By applying the Precise Timing Spike Detection
algorithm~\cite{PTSD} in the BrainWave software to the recorded
time series $y_i(t)$ of the $i$-th electrode, we obtain the times
of the spikes $t^{(i)}_k$, $k=1, 2, \ldots, N_i$, for the $N_i$
spikes detected for node $i$. The spike train $S_i(t)$ is
constructed as follows: $S_i(t)=1$ for $t=t^{(i)}_k$, $k=1, 2,
\ldots, N_i$ and $0$ otherwise. The general idea is to estimate a
functional link between nodes $i$ and $j$ when the correlation
between the two spike trains exceeds a certain threshold. The
cross-correlation of the spiking activity of nodes $i$ and $j$ is
measured by $C_{ij} \in [0,1]$, which is defined
by~\cite{Maccione2012} {\blue \begin{equation} C_{ij}(\tau) =
\frac{1}{\sqrt{N_i N_j}} \sum_{k=1}^{N_i}
S_i(t^{(i)}_k)S_j(t^{(i)}_k+\tau) \label{Cijtau}
\end{equation}}
In the earlier method~\cite{Maccione2012}, $C_{ij}(\tau)$ is
calculated in a window of $\tau$-values: $-n \Delta \le \tau \le n
\Delta$, where $\Delta$ is the sampling time interval and if the
maximum value $C_{ij}(\tau_0)$ at $\tau_0$ within this window
exceeds a certain threshold then a link of weight $C_{ij}(\tau_0)$
is inferred between nodes $i$ and $j$ with the direction of the
link determined by the sign of $\tau_0$. To detect also inhibitory
links with negative weights, $f_{ij}(\tau)$ measuring the
difference of $C_{ij}(\tau)$ from its average value in the window
has been introduced~\cite{Pastore2018}
\begin{equation} f_{ij}(\tau) = C_{ij}(\tau)-
\frac{1}{2n+1}\sum_{k=-n}^n C_{ij}(k \Delta), \label{fij}
\end{equation} which
can assume both positive and negative values. In our calculations,
we use $n=100$. Let $|f_{ij}(\tau)|$ attain its maximum at
$\tau=\tau^*$~\cite{note}; a link is inferred for the functional
connectivity, from node $i$ to node $j$ if $\tau^*>0$ and from
node $j$ to node $i$ if $\tau^*<0$, with strength $f_{ij}(\tau^*)$
when $|f_{ij}(\tau^*)|$ exceeds a certain threshold and $|\tau^*|$
is not shorter than the time needed for a synaptic signal with a
propagation speed of 400mm/s~\cite{Pastore2018}. Excitatory links
have $f_{ij}(\tau^*)>0$ while inhibitory links have
$f_{ij}(\tau^*)<0$. The requirement that a node is either
excitatory or inhibitory, with outgoing links having only one
sign, cannot be enforced in this cross-correlation based method
for estimating functional connectivity. This method also has a
considerably weaker sensitivity for detecting inhibitory
links~\cite{Pastore2018}. In contrast, the covariance-relation
based method that we use to estimate effective connectivity can
detect excitatory and inhibitory links equally well using only a
recording time of 5 minutes.

\begin{figure}[htbp]
                    \centering
\includegraphics[width=2.5in]{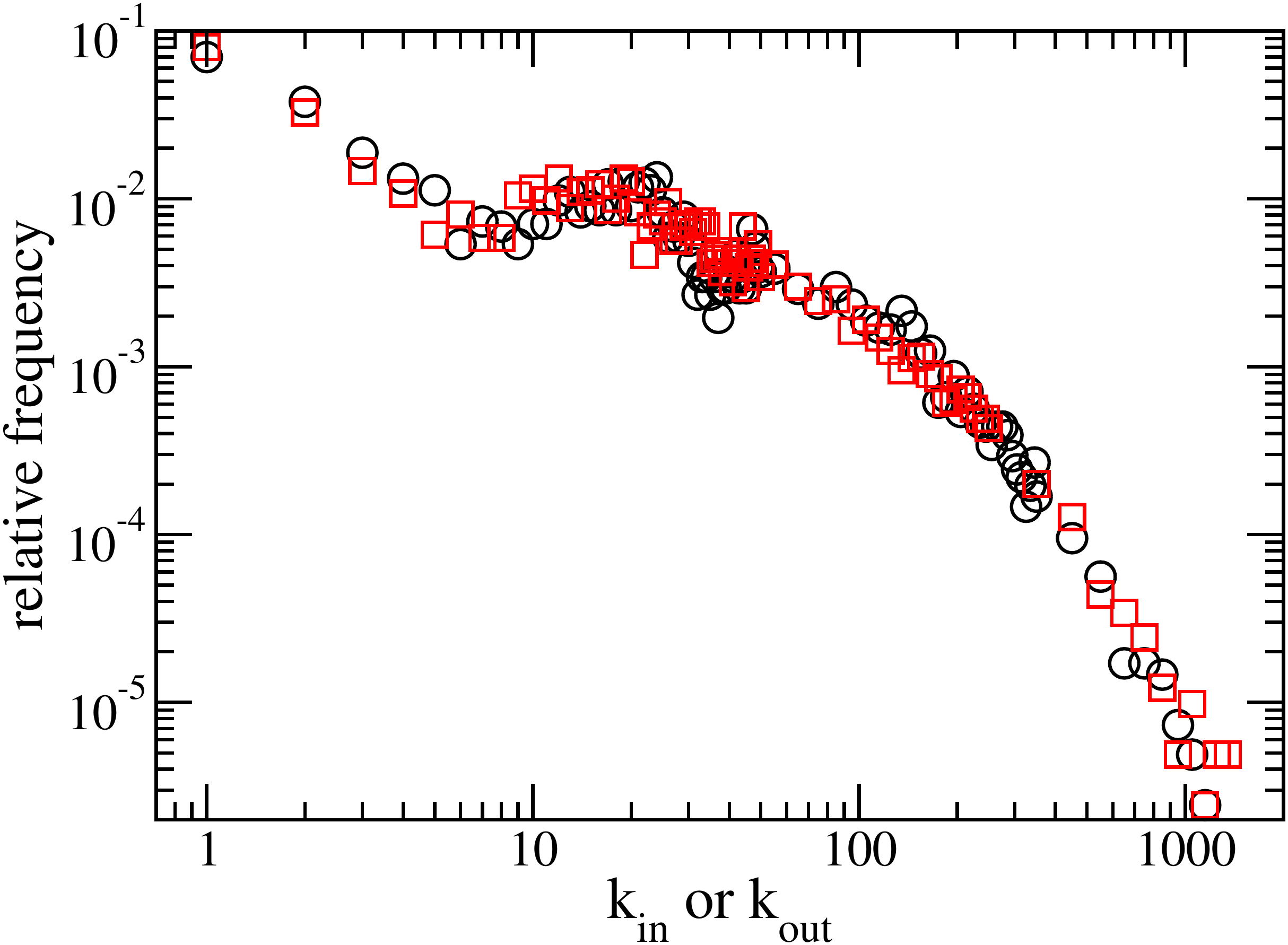}
\caption{Distributions of the incoming and outgoing degree,
$k_{\rm in}$ (circles) and $k_{\rm out}$ (squares) of the
 functional connectivity estimated for case 3.}
                    \label{fig6}
                    \end{figure}

\begin{figure}[htbp]
\centering
\includegraphics[width=1.2in]{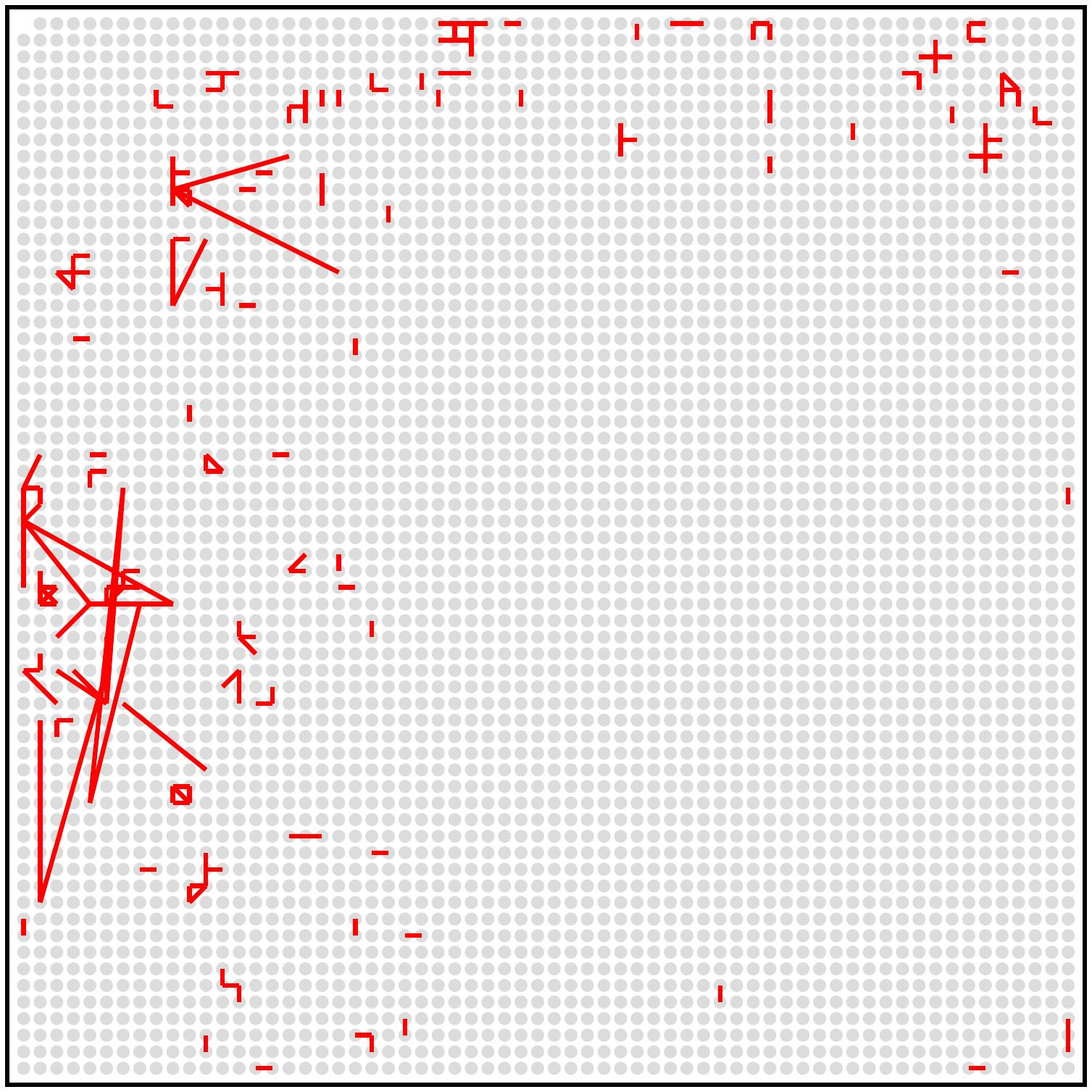}
\includegraphics[width=1.2in]{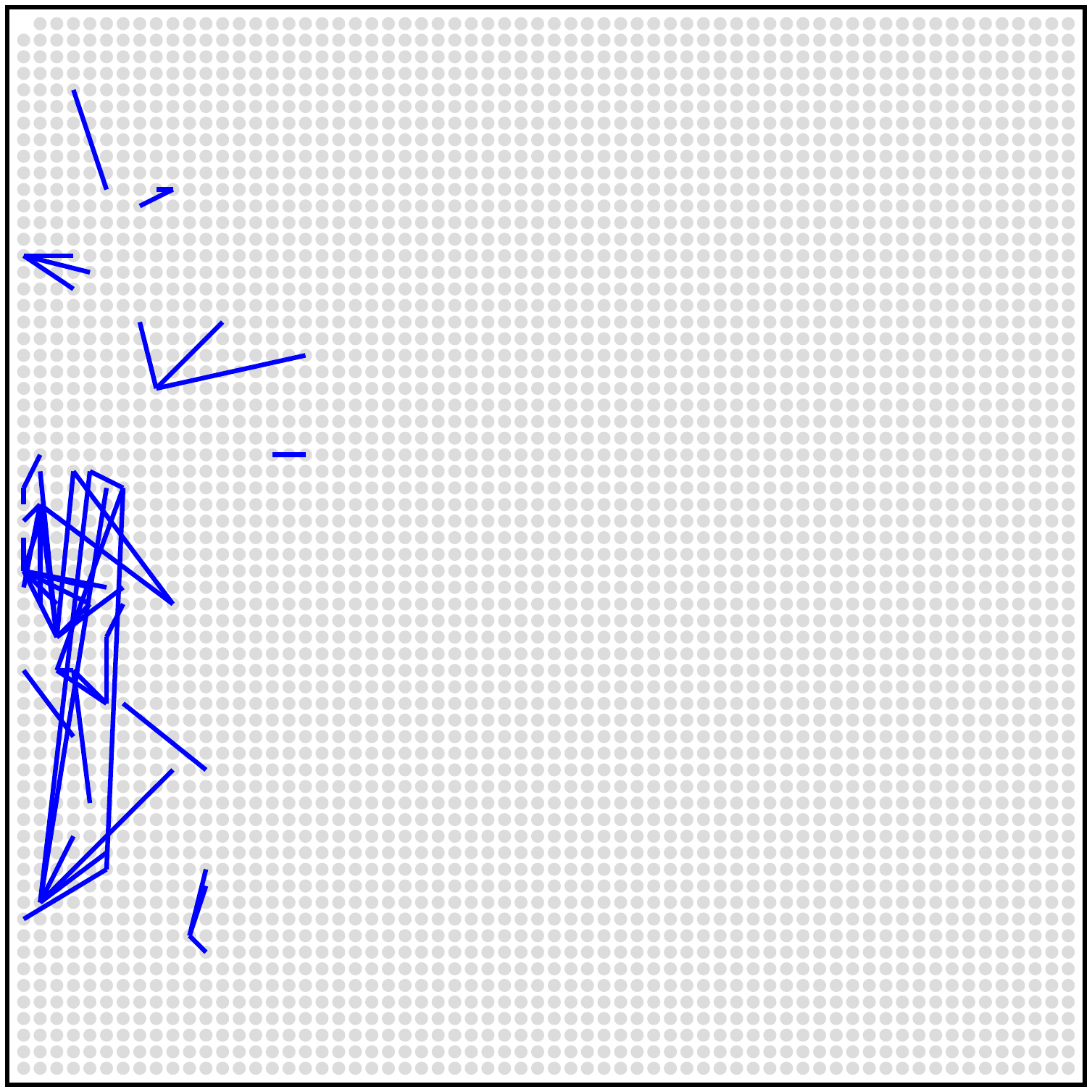}
\includegraphics[width=1.2in]{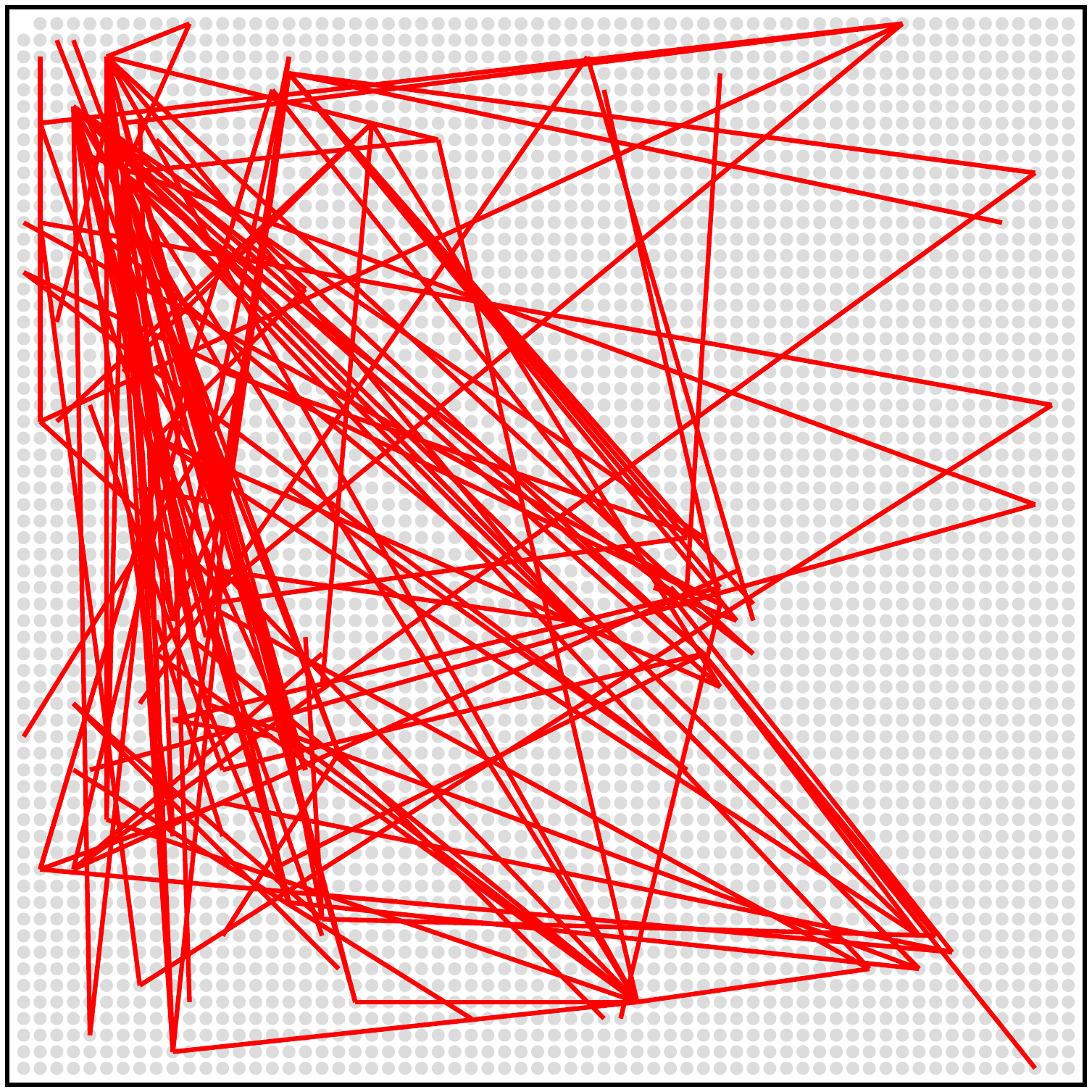}
\includegraphics[width=1.2in]{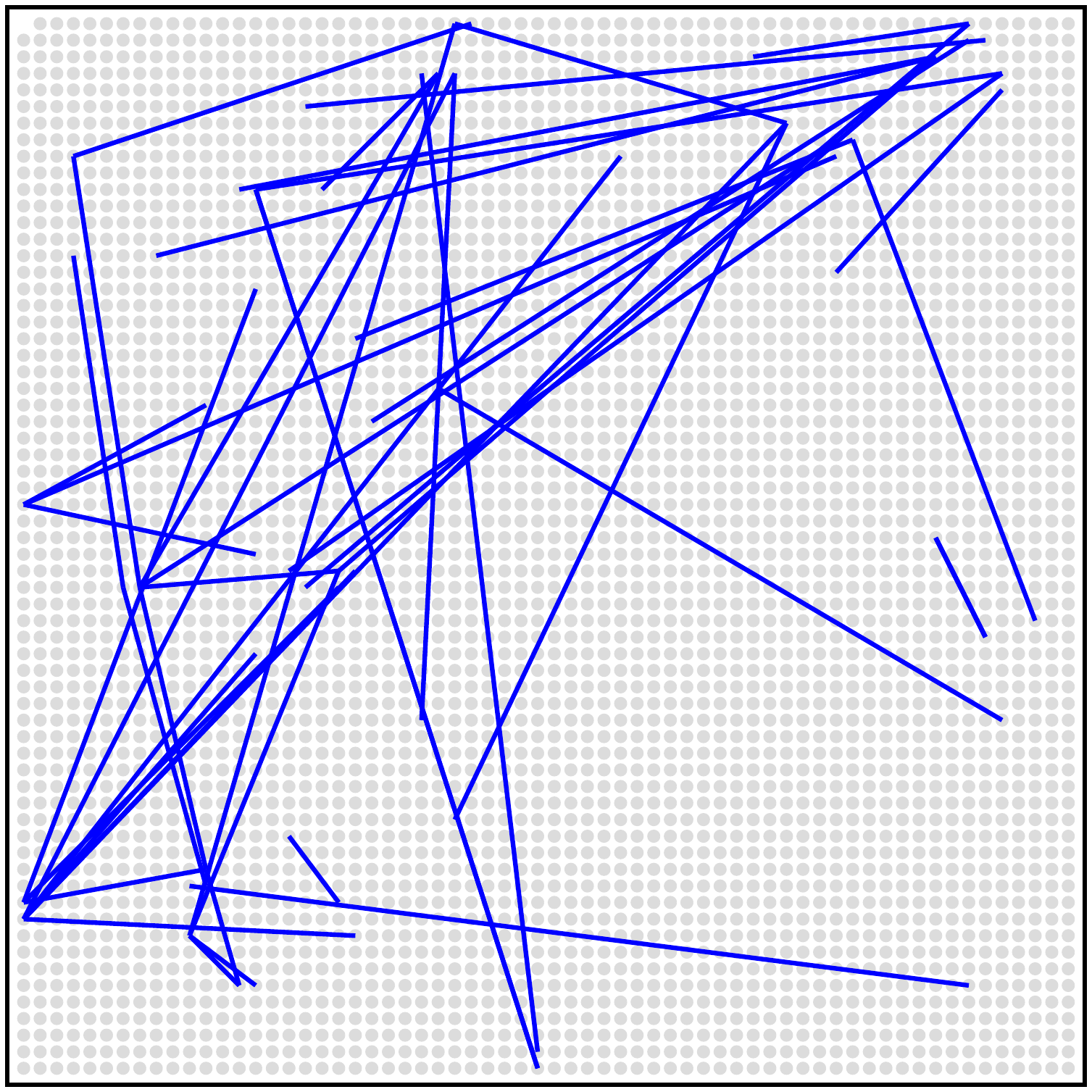}
\caption{Comparison of the strongest $N_{\rm exc}$ excitatory
(red) and the strongest $N_{\rm inh}$ inhibitory links (blue) of
the effective connectivity (top panels) and of the functional
connectivity (bottom panels) for case 3. $N_{\rm inh}=50$ for both
types of connectivity, $N_{\rm exc}=200$ for effective
connectivity and $N_{\rm exc}=151$ for functional connectivity.}
            \label{fig7}
\end{figure}

{\blue Comparing to effective connectivity, a large fraction (0.2
to 0.46) of nodes
 have zero incoming  or outgoing
degrees} in the estimated functional connectivity. In
Fig.~\ref{fig6}, we show the distributions of nonzero incoming and
outgoing degrees for functional connectivity. {\BLUE It can be
seen that the two distributions are similar and are clearly
different than the incoming and outgoing degree distributions
found in effective connectivity~(see Fig.~\ref{fig2}). The degree
distribution in functional connectivity estimated from
measurements recorded in a much longer duration of one hour has
been found to be scale-free~\cite{Pastore2018}}. We compare the
strongest excitatory and inhibitory links estimated in the two
types of connectivity in Fig.~\ref{fig7}. We choose 200 strongest
excitatory links and 50 strongest inhibitory links in the
effective connectivity. It is not possible to choose exactly the
same number of links in functional connectivity as many links have
the same coupling strength and we choose the closest number of
links for the comparison. In the effective connectivity, most of
the strongest excitatory links are connecting nearby nodes. This
feature is not found in the functional connectivity.

\begin{figure}[htbp]
\centering
\includegraphics[width=2.5in]{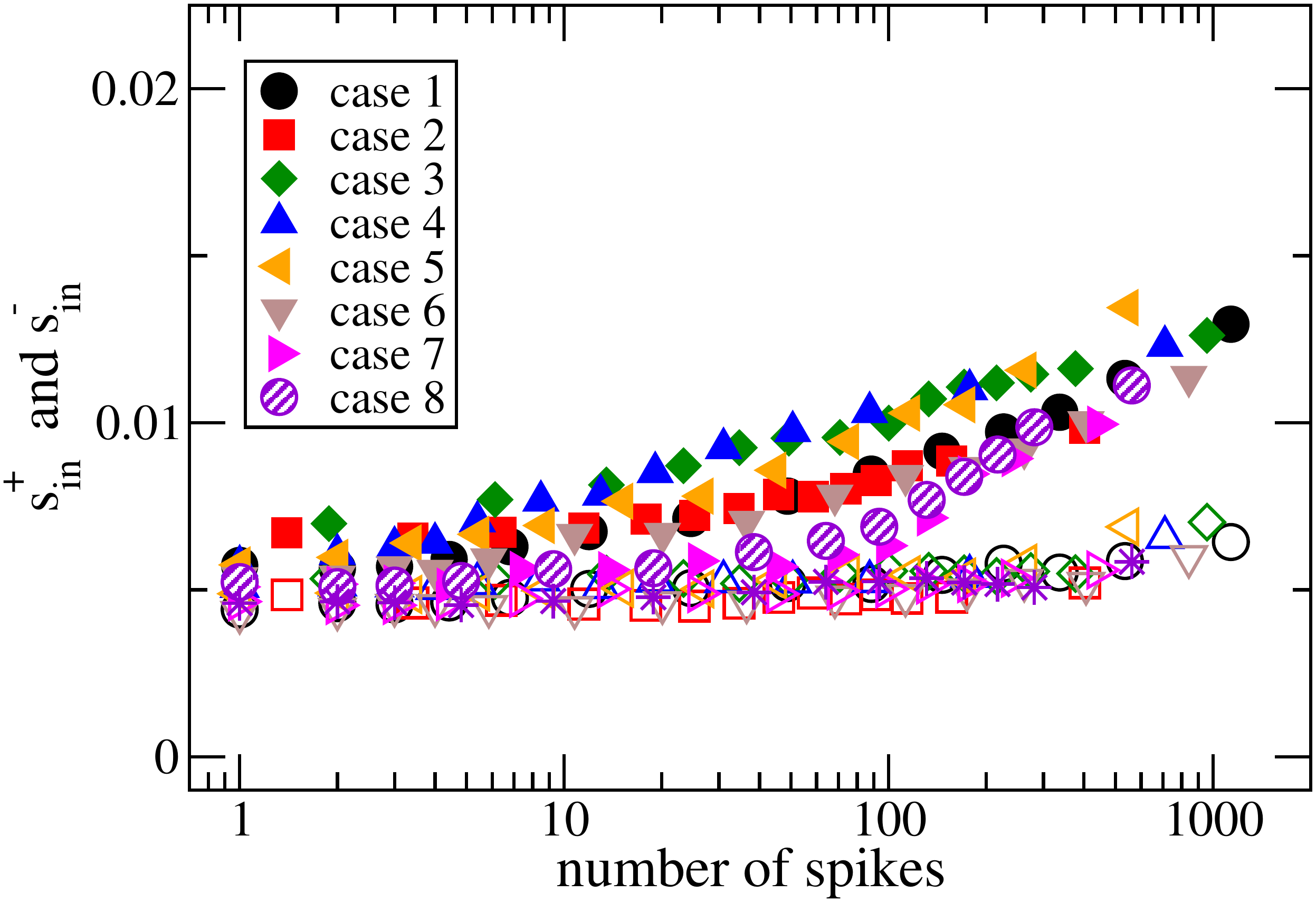}
\includegraphics[width=2.5in]{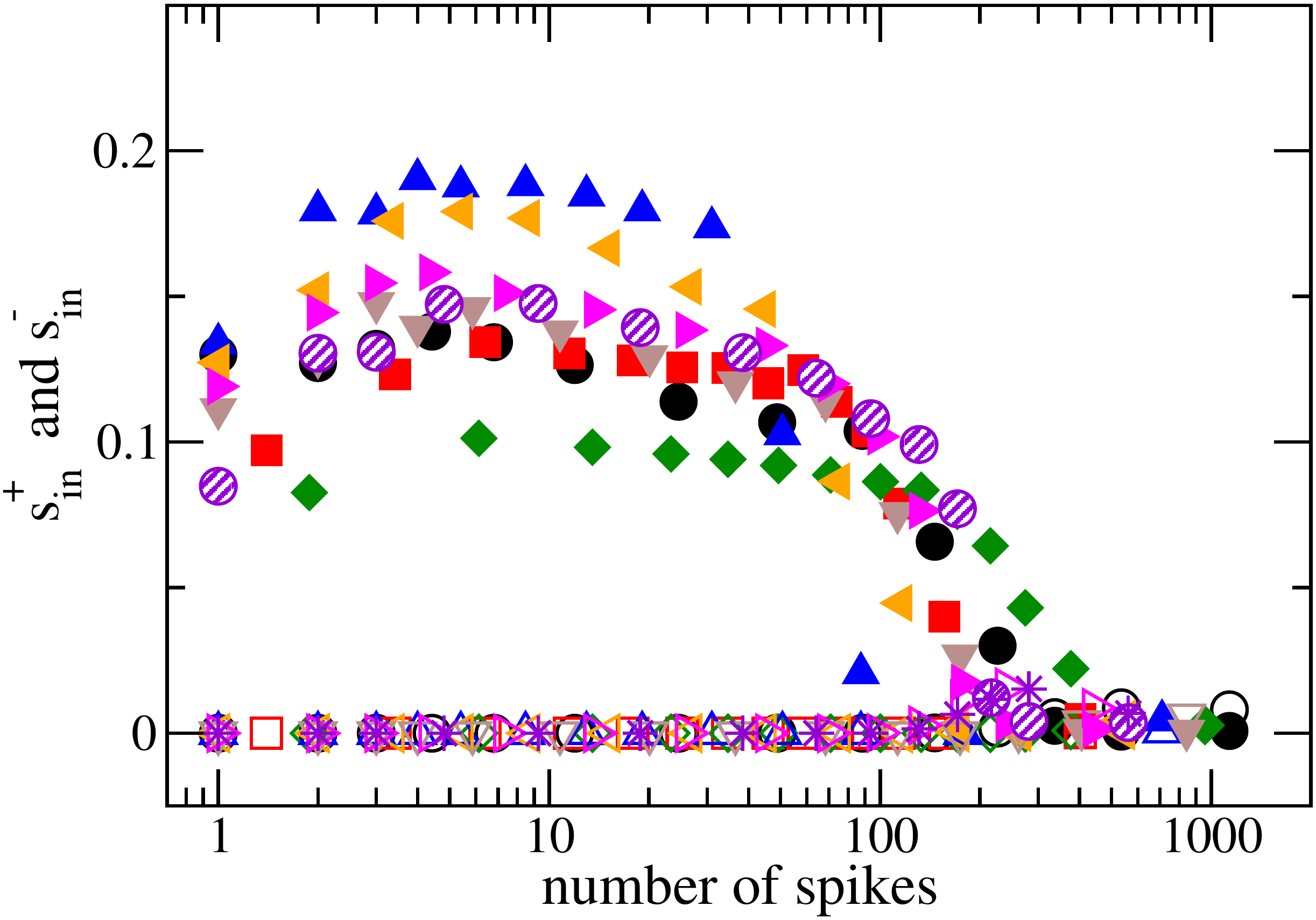}
\caption{The dependence of the mean values of $s_{\rm in}^+$ and
$s_{\rm in}^-$ on the number of detected spikes of the nodes in
effective connectivity (top panel) and functional connectivity
(bottom panel). The filled and shaded symbols shown in the legend
denote $s_{\rm in}^+$ and $s_{\rm in}^-$ are denoted by the
corresponding open symbols for cases 1-7 and by the star symbols
for case 8.}
            \label{fig9}
\end{figure}

\begin{figure}[htbp]
\centering
\includegraphics[width=2.5in]{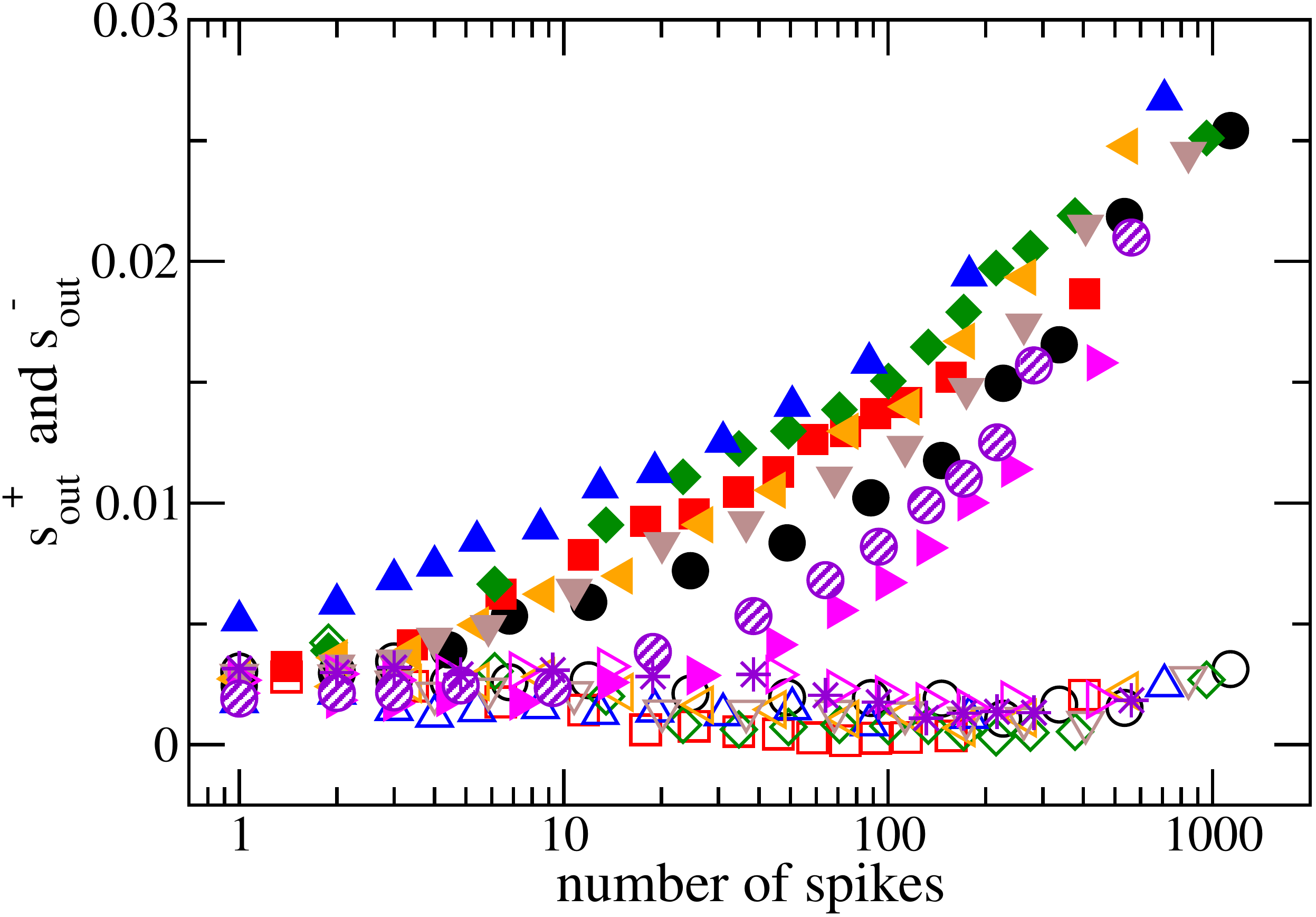}
\includegraphics[width=2.5in]{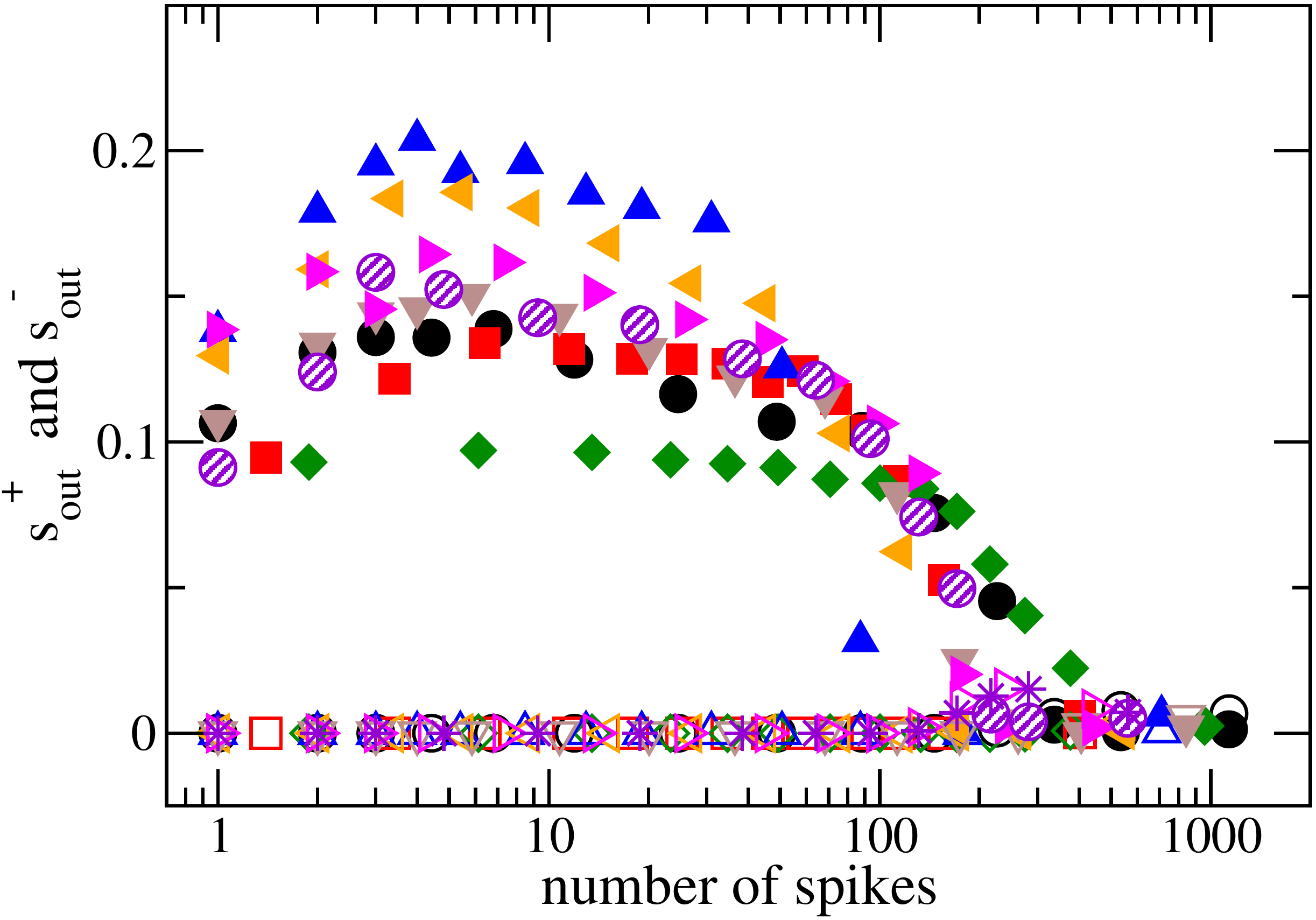}
\caption{The dependence of the mean values of $s_{\rm out}^+$ and
$s_{\rm out}^-$ on the number of detected spikes of the nodes in
effective connectivity (top panel) and functional connectivity
(bottom panel). Same symbols as in Fig.~\ref{fig9}}.
            \label{fig10}
\end{figure}

To check whether the proposed effective connectivity can indeed
reveal relationships between network structure and dynamics better
than functional connectivity, we study the relation between the
spiking activity and the average synaptic strength of the nodes.
Specifically, we divide the nodes into several groups according to
their number of detected spikes and calculate the mean values of
the average synaptic strength of excitatory and inhibitory
incoming and outgoing links of these different groups. Since nodes
have both excitatory and inhibitory outgoing links in the
estimated functional connectivity, we generalize the definition of
$s_{\rm out}$ to $s_{\rm out}^+$ and $s_{\rm out}^-$:
\begin{eqnarray}
s_{\rm out}^+(i) &\equiv& \frac{\sum_{j\ne i, w_{ji}>0} w_{ji}}{k_{\rm out}^+(i)} , \label{soutplus} \\
 s_{\rm out}^-(i) &\equiv& \frac{\sum_{j\ne i, w_{ji}<0} w_{ji}}{k_{\rm out}^-(i)} , \label{soutminus}
 \end{eqnarray}
{\BLUE and $s_{\rm out}^+$ ($s_{\rm out}^-$) is equal to zero for
nodes without excitatory (inhibitory) outgoing links.} For the
effective connectivity, $s_{\rm out}^+$ ($s_{\rm out}^-$) is equal
to $s_{\rm out}$ for excitatory (inhibitory) nodes. The results
for the two types of connectivity are shown in Figs.~\ref{fig9}
and \ref{fig10}. {\BLUE The dependence of the mean values of
$s_{\rm in}^-$ and $s_{\rm out}^-$ on the spiking activity is weak
for both types of connectivity. Definite dependencies of the mean
values of $s_{\rm in}^+$ and $s_{\rm out}^+$ on the number of
detected spikes are found in our estimated effective connectivity
but not in the estimated functional connectivity.} In particular,
the intuition that
 nodes with larger $s_{\rm in}^+$
would spike more is supported by the effective connectivity but
not functional connectivity. The effective connectivity further
reveals that nodes spike more have larger $s_{\rm out}^+$ on
average and such a definite dependence is again lacking in the
functional connectivity.

\section{Summary and Conclusions} \label{Summary}

Revealing connectivity of neuronal networks from measurements
taken by large-scale MEA is a challenging inverse problem.
Existing methods focus on functional connectivity that is based on
the statistical correlation of the detected spiking activities.
{\BLUE As statistical correlation arising from indirect influences
lead to false positives, efforts have been spent to optimize the
performance of inference based on spike cross-correlation. In a
recent study~\cite{Kobayashi2019}, a method, {\Blue denoted as
GLMCC,} has been developed by applying a generalized linear model
to spike cross-correlations which shows how an increase of the
duration of spike recording can improve the performance of
inference. This study further gives an analytical estimate of the
required duration for reliable inference which is inversely
related to the firing rates of the neurons and the synaptic
strength of the connections. In our work, we study effective
connectivity, which measures direct interactions and should be
more relevant and useful for studying the relationships between
network structure and dynamics and between network structure and
functions of neuronal networks.} We have  modelled in vitro
neuronal cultures as stochastic dynamical systems and adopted a
general method that reconstructs {\blue directed links and their
relative weights of a network} from
dynamics~\cite{ChingTamPRE2017} to estimate effective connectivity
of neuronal cultures from voltage measurements recorded by MEA.
{\Blue This general method makes use of Eq.~(\ref{relation2}),
which follows from Eq.~(\ref{relation}), a relation between the
time-lagged covariance and the equal-time covariance of the
dynamics to extract information of the direct interactions of the
nodes. Equation~(\ref{relation}) is derived for systems that
approach a fixed point in the noise-free limit and numerical
results revealed that Eq.~(\ref{relation2}) also holds
approximately for systems that fluctuate around oscillatory
FitzHugh-Nagumo dynamics or chaotic Rossler
dynamics~\cite{PhysicaA2018,ChingTamPRE2017,TamPhDthesis} even
though Eq.~(\ref{relation}) cannot be derived. Motivated by these
numerical results, we assume Eq.~(\ref{relation2}) to hold for our
model. For networks of stochastic binary neurons with activity
being either 0 or 1, a relation between the time-lagged covariance
and equal-time covariance similar to Eq.~(\ref{relation}) can be
obtained~\cite{RoudiPRL2011,Mezard2011,Frontiers2013,PRX2016} and
used to reconstruct the effective synaptic couplings of the
neurons~\cite{PRX2016}. In our model, the activity of each neuron
is described by the continuous variable $x_i(t)$ and the
time-lagged and equal-time covariances are calculated using the
whole time series of the voltage measurements and not only the
spike trains of 0's and 1's.} When applying this method to
reconstruct real network from experimental measurements, we have
to overcome additional difficulties. One major difficulty is that
this involves a calculation of a principal matrix logarithm, which
is very sensitive to noise in the data. If the method is applied
directly to the MEA voltage recordings, a complex matrix would be
obtained. By first applying a moving average filter to the MEA
voltage recordings to reduce the effect of noise, the problem of
complex matrix has been avoided and we have successfully estimated
the effective connectivity, {\blue namely the directed
interactions with their synaptic weights,} of neuronal networks of
over 4000 nodes from relatively short MEA recordings of 5 min.
{\Blue In comparison, good performance of the GLMCC method has
been shown using numerical data with a much longer duration of 90
min of spike recording~\cite{Kobayashi2019}. Moreover,} as our
method makes use of the whole voltage time series and not only the
detected spikes, low firing rates in the MEA recording do not pose
an additional difficulty.

Our results of the effective connectivity reproduce various
reported features of cortical regions in rats and monkeys and has
similar network properties as the nematode C. elegans chemical
synaptic network, {\Blue supporting that the estimated effective
connectivity can capture the general properties of synaptic
connections.} {\BLUE Moreover, numerical simulations of networks
of spiking neurons using the estimated effective connectivity
reproduce the long-tailed distribution of firing rates found in
the MEA recordings~\cite{Frontiers}.} The distributions of
incoming and outgoing degrees are different from the reported
scale-free distributions in the functional
connectivity~\cite{Pastore2018}. In particular the excitatory and
inhibitory incoming degree distributions are found to be bimodal.
There have been studies indicating that the robustness of
undirected networks against both random failures and targeted
attacks can be optimized by having a bimodal degree
distribution~\cite{Stone2004,Stanley2005} and future studies are
required to establish the significance of the bimodal feature of
incoming degree distributions. We have found that the
distributions of average synaptic strengths are non-Gaussian and
skewed with a long tail and that the distribution of the average
synaptic strength of outgoing links for excitatory nodes with
positive average synaptic strength of incoming links ($s_{\rm
in}>0$) is approximately lognormal, and our results are consistent
with reported results found in in vitro and in vivo studies of
synaptic strengths in the cortex~\cite{Buzsaki2014}. The
significance of such non-Gaussian distributions with long tails is
that a small fraction of nodes have dominantly strong average
synaptic strength suggesting the possibility that the bulk of the
information flow occurring mostly through them~\cite{Buzsaki2014}.
{\blue It would be interesting to understand how the long-tailed
synaptic strength distribution might be related to the spiking and
bursting dynamics.}

The effective connectivity and functional connectivity estimated
from the same sets of MEA recordings are different. {\BLUE The
average synaptic strengths of excitatory incoming and outgoing
links are found to increase with the spiking activity in the
estimated effective connectivity but not in the estimated
functional connectivity.} These findings demonstrate that the
effective connectivity estimated in the present work can indeed
better reveal relationships between network structure and dynamics
for neuronal cultures. {\blue Understanding the relationships
between network structure and dynamics will be a topic of interest
in future studies.}

\vspace{1cm}

\begin{acknowledgments} The work of CS, KCL, CYY and ESCC has been
supported by the Hong Kong Research Grants Council under grant no.
CUHK 14304017. PYL was supported by the Ministry of Science and
Technology of the Republic of China under grant no.
107-2112-M-008-013-MY3 and NCTS of Taiwan.
\end{acknowledgments}

\vspace{0.5cm}

\appendix*

\vspace{0.2cm}

\centerline{Appendix: Experimental details on the neuronal
culture}

\vspace{0.25cm}

The complementary metal-oxide-semiconductor (CMOS)-based high
density multielectrode array (HD-MEA) was pre-coated with 0.1 \%
Poly-D-lysine (Sigma P6407) and 0.1 \% adhesion proteins laminin
(Sigma L2020). After plating on the HD-MEA chip, cultures were
filled with 1~ml of culture medium~[DMEM (Gibco 10569) + 5\% FBS
(Gibco 26140) + 5\% HS (Gibco 16050) + 1\% PS (Gibco 15140)] and
placed in a humidified incubator (5\% CO$_{2}$, 37$^{\circ}$C).
Half of the medium was replaced by Neurobasal medium supplemented
with B27 [Neurobasel medium (Gibco. 21103) + 2\% 50X B27
supplement (Gibco. 17504) + 200 $\mu M$ GlutaMAX (Gibco 35050)]
twice a week.


\begin{thebibliography}{10}
\bibitem{note} When
$|f_{ij}(\tau)|$ attains its maximum at more than one value of
$\tau$, all of the same sign, $\tau^*$ is taken as the
$\tau$-value with the minimum magnitude. If the values of $\tau$
at which $|f_{ij}(\tau)|$ attains its maximum include both
positive and negative values, then we take $\tau^*$ as the
$\tau$-value with the minimum magnitude separately for the
positive and negative $\tau$-values. If these two $\tau^*$ both
satisfy the criterion set by the maximum propagation speed of
synaptic signal, then there are bidirectional links between nodes
$i$ and $j$ with the same coupling strength in either direction.
\end{thebibliography}
\end{document}